\documentclass[aip,pop,preprint,nofootinbib]{revtex4-1}
\usepackage{amsfonts}
\usepackage{amsmath}
\usepackage{graphicx}
\usepackage{color}

\newcommand{\curl}{\operatorname{curl}}
\newcommand{\ddt}[0]{\frac{d}{dt}}

\newcommand{\divergence}{\operatorname{div}}

\newcommand{\exteriord}{\mathrm{d}} 
\newcommand{\grad}{\operatorname{grad}}
\newcommand{\half}{\frac{1}{2}}
\newcommand{\norm}[1]{\| #1 \|}
\newcommand{\order}[1]{\mathcal{O}(#1)}
\newcommand{\partiald}[2]{\frac{\partial #1}{\partial #2}}

\newcommand{\reals}{\mathbb{R}} 

\newcommand{\totald}[2]{\frac{d #1}{d #2}}

\newcommand{\vectorTwo}[2]{\left(\begin{array}{c}#1 \\ #2 \end{array} \right)}

\newcommand{\matrixTwobyTwo}[4]{\left(\begin{array}{cc} #1 & #2 \\ #3 & #4 \end{array} \right)}

\makeatletter
\renewcommand\d[1]{\mspace{6mu}\mathrm{d}#1\@ifnextchar\d{\mspace{-3mu}}{}}
\makeatother

\newcommand{\myepigraph}[3]{
  \setlength\epigraphwidth{14 cm}
  \setlength\epigraphrule{0pt}
  \epigraph{\textit{#1 \newline \vspace{-0.1in}}}{---\ifx&#3&#2 \else #2, \fi
  \textit{#3}}
  \vspace{0.2 in}
}

\makeatletter
\g@addto@macro\@floatboxreset\centering
\makeatother

\makeatletter
\@ifundefined{theorem}{

}{}
\makeatother

\makeatletter
\@ifundefined{proof}{

}{}
\makeatother

\makeatletter
\@ifundefined{definition}{

}{}
\makeatother

\makeatletter
\@ifundefined{example}{

}{}
\makeatother

\makeatletter
\@ifundefined{qed}{
\newcommand{\qed}{\nobreak \ifvmode \relax \else
      \ifdim\lastskip<1.5em \hskip-\lastskip
      \hskip1.5em plus0em minus0.5em \fi \nobreak
      \vrule height0.75em width0.5em depth0.25em\fi}
		}{}
\makeatother

\begin{document}

\title{Degenerate Variational Integrators for Magnetic Field Line Flow and Guiding Center Trajectories}
\author{C. L. Ellison}
\email{ellison6@llnl.gov}
\affiliation{Lawrence Livermore National Laboratory, Livermore, CA 94550, USA}
\affiliation{Princeton Plasma Physics Laboratory, Princeton, NJ 08543, USA}
\author{J. M. Finn}
\altaffiliation[Present address: ]{Tibbar Plasma Technologies, 274 DP Rd, Los Alamos, NM 87544, USA}
\affiliation{Los Alamos National Laboratory, Los Alamos, NM 87545, USA}
\author{J. W. Burby}
\affiliation{Courant Institute of Mathematical Sciences, New York, New York 10012, USA}
\author{M. Kraus}
\affiliation{Max-Planck-Institut f\"ur Plasmaphysik, Garching, Deutchland}
\author{H. Qin}
\affiliation{Princeton Plasma Physics Laboratory, Princeton, NJ 08543, USA}
\affiliation{Department of Modern Physics, University of Science and Technology of China, Hefei, Anhui 230026, China}
\author{W. M. Tang}
\affiliation{Princeton Plasma Physics Laboratory, Princeton, NJ 08543, USA}
\date{\today}

\begin{abstract}
Symplectic integrators offer many advantages for the numerical solution of Hamiltonian differential equations, including bounded energy error and the preservation of invariant sets. Two of the central Hamiltonian systems encountered in plasma physics --- the flow of magnetic field lines and the guiding center motion of magnetized charged particles --- resist symplectic integration by conventional means because the dynamics are most naturally formulated in non-canonical coordinates, i.e., coordinates lacking the familiar $(q, p)$ partitioning. Recent efforts made progress toward non-canonical symplectic integration of these systems by appealing to the variational integration framework; however, those integrators were multistep methods and later found to be numerically unstable due to parasitic mode instabilities. This work eliminates the multistep character and, therefore, the parasitic mode instabilities via an adaptation of the variational integration formalism that we deem ``degenerate variational integration''. Both the magnetic field line and guiding center Lagrangians are degenerate in the sense that their resultant Euler-Lagrange equations are systems of first-order ODEs. We show that retaining the same degree of degeneracy when constructing a discrete Lagrangian yields one-step variational integrators preserving a non-canonical symplectic structure on the original Hamiltonian phase space. The advantages of the new algorithms are demonstrated via numerical examples, demonstrating superior stability compared to existing variational integrators for these systems and superior qualitative behavior compared to non-conservative algorithms.
\end{abstract}

\maketitle

\section{Introduction}
\label{sec:introduction}

Two of the foundational dynamical systems describing magnetized plasmas, the flow of magnetic field lines and the motion of guiding center trajectories, have long been known to be Hamiltonian\cite{Littlejohn_1983, Cary_1983, Cary_2009}. The Hamiltonian character of these systems enables the use of powerful analytic tools: the famed KAM theorem bounds the area of stochastic regions in resonantly perturbed tokamaks, for example \cite{Landau_Lifshitz_1969, Lichtenberg_1983, Marsden_1999}. The Hamiltonian character is equally important for computationally modeling magnetized plasmas: the Liouville theorem is critical for particle-based methods, such as drift- and gyro-kinetic simulations, in which simulated particles advect volume elements of the distribution function, for instance. Outside of plasma physics, the numerical integration of Hamiltonian systems has also benefitted from powerful numerical methods known as symplectic integrators \cite{Hairer_2006_symplectic, McLachlan_2006}. Symplectic integrators possess an area-preserving property that allows them to retain the Hamiltonian character in numerical trajectories and thereby obtain excellent long term fidelity. They have proven indispensable for orbital mechanics and particle accelerators \cite{Forest_1990, McLachlan_2006}, for example, and are strong candidates for improved numerical methods in plasma physics \cite{Qin_2008, Qin_2009, Squire_2012_PIC, Qin_2016}.

Unfortunately, existing symplectic integrators cannot be readily applied to magnetic field line flow or guiding center trajectories. Conventional symplectic integrators are formulated in terms of \emph{canonical} coordinates, that is, coordinates that may be partitioned into positions $q$ and conjugate momenta $p$ whose dynamics are governed by Hamilton's equations in canonical form. Magnetic field line flow and guiding center trajectories are Hamiltonian in a more general sense; the most natural coordinates admit no such partitioning, and are thus deemed \emph{non-canonical} coordinates. Although one may transform these systems to canonical coordinates, at least locally \cite{Darboux_1882}, the reliance on such a transformation in a numerical scheme incurs computational overhead, decreasing the advantage of using otherwise powerful symplectic integrators. The development of symplectic integrators for non-canonical Hamiltonian systems remains an outstanding challenge in numerical analysis \cite{Karasozen_2004}. In the meantime, magnetized plasma simulations involving the advance of guiding center trajectories resort to non-symplectic algorithms \cite{White_1984, Velasco_2012, Kramer_2013, Pfefferle_2014, Hirvijoki_2014, Hirvijoki_2015}.

Recently, promising progress toward symplectic integration of guiding center trajectories has been made by applying the theory of variational integration \cite{Qin_2008, Qin_2009, Li_2011, Kraus_2013}. Instead of applying discrete-time approximations directly to the equations of motion, ``variational integrators'' were constructed by discretizing time in the Lagrangian that underlies the equations of motion by way of a variational principle \cite{Marsden_2001}. By introducing all truncation error into the Lagrangian, the algorithms are guaranteed to preserve \emph{some} symplectic two-form \cite{Marsden_2001}. Although initial results in the guiding center context exhibited the desired long-term numerical fidelity, additional testing revealed numerical instabilities that rendered them unfit for widespread use \cite{Squire_2012, Ellison_2015_PPCF}. The cause of the instabilities was traced \cite{Ellison_2015_PPCF, Ellison_thesis} to the fact that the variational integrators were \emph{multistep} methods; the discrete equations were of higher order than the continuum equations, requiring additional initial conditions, preserving areas in a higher-dimensional phase space, and introducing unphysical ``parasitic modes''. Before variational integrators can be considered to be robust methods for magnetic field line flow and guiding center trajectories, these instabilities must be eliminated.

In this contribution, stable variational integrators are constructed for magnetic field line flow and guiding center trajectories through a novel approach. Specifically, emphasis is placed on retaining a property exhibited by the magnetic field line and guiding center Lagrangians known as \emph{degeneracy}. These Lagrangians are degenerate in the sense that their corresponding Euler-Lagrange equations are systems of first-order --- rather than second-order --- ODEs. This is because these Lagrangians are examples of so-called ``phase-space Lagrangians'' \cite{Goldstein_2001, Cary_1983}, Lagrangians whose Euler-Lagrange equations are Hamilton's equations, i.e., a system of first-order ODEs. The degeneracy of phase-space Lagrangians is only problematic when attempting to formulate variational integrators, which are conventionally assumed to be formed from non-degenerate Lagrangians \cite{Marsden_2001}. The instabilities in previous variational guiding center integrators may be attributed to the fact that degeneracy was \emph{lost} during the discretization procedure, resulting in a system of second-order difference equations, as would be appropriate for modeling a system of second-order differential equations. By retaining the degeneracy, the variational integrators obtained here are instead one-step methods, requiring only a single initial condition and preserving non-canonical symplectic structures in the original Hamiltonian phase space. We deem the new integrators ``degenerate variational integrators'' (DVIs) to emphasize the importance of this property. When variational integrators for degenerate Lagrangians have been studied in the past, degeneracy of the variational integrators was avoided \cite{Rowley_2002, Ober-Blobaum_2013, Tyranowski_2014}; here we advocate degeneracy of the integrator as beneficial for the stable integration of degenerate Lagrangian systems. We provide a simple and helpful condition for checking whether a particular discretization is degenerate.

For the magnetic field line DVI, the only restriction used to obtain the desired degree of degeneracy is the choice of an electromagnetic gauge wherein one component of the magnetic vector potential is zero; the fields are otherwise arbitrary. For the guiding center DVI, one component of the magnetic vector potential is set to zero and it is further assumed that the same covariant component of the magnetic field is zero (in the chosen coordinates). Although many applications of interest do not satisfy this property, this simplification enabled the present progress en route to completing the general problem. In another publication \cite{Burby_2017}, this restriction is avoided through a re-definition of the guiding center coordinates. The new coordinates are ``regularized'', eliminating the large parallel velocity singularity from the guiding center equations. Simultaneously, the new coordinates enable construction of a guiding center DVI (using the techniques presented in this paper) without any restrictions beyond the existence of a non-vanishing toroidal component of the magnetic field, assumed for the regularizing transformation. For the scope of the present work, we emphasize the DVI technique and restrict attention to the conventional guiding center Lagrangian \cite{Littlejohn_1983} subject to the aforementioned condition on the coordinates.

The structure of the manuscript is as follows. The challenges of variational integration of degenerate Lagrangian systems and the proposed solution are described in Section \ref{sec:dvi_method}. To begin in a familiar setting, Section~\ref{ssec:canonical_hamiltonian_systems} illustrates the procedure for canonical Hamiltonian systems, recovering a well-known symplectic integrator as a degenerate variational integrator. Section \ref{ssec:magnetic_field_line_flow} then reviews the non-canonical Hamiltonian description of magnetic field line flow and derives a DVI for these dynamics. In Section \ref{ssec:guiding_center_trajectories}, a guiding center DVI is developed, and the tradeoffs relative to canonical symplectic integration \cite{Zhang_2014} or projected variational integrators \cite{Kraus_2017_PVI} are discussed. Section \ref{sec:numerical_demonstrations} numerically demonstrates the benefits of the non-canonical DVIs, including the elimination of the numerical instabilities present in previous variational integrators and superior qualitative behavior to commonly used Runge-Kutta schemes. Concluding remarks are presented in Section \ref{sec:discussion}. Additionally, a brief introduction to differential geometry is presented in Appendix~\ref{sec:differential_geometry_primer} to facilitate understanding of the notation used in the main sections, and Appendix~\ref{sec:fermats_principle} discusses Fermat's Principle as a simple example of a system with a degenerate Lagrangian.

\section{Degenerate Variational Integrators}
\label{sec:dvi_method}

\subsection{Canonical Hamiltonian Systems}
\label{ssec:canonical_hamiltonian_systems}

To (i) demonstrate the need for and (ii) illustrate the development of degenerate variational integrators, we will begin in the familiar context of canonical Hamiltonian systems. Although a plethora of well-established techniques exist for deriving symplectic integrators in canonical coordinates \cite{Ruth_1983, Channell_1990, Forest_1990, Sanz-Serna_1988, Marsden_2001, Hairer_2006_symplectic, Leok_2011}, this subsection introduces the key terminology, issues, and methods to be employed in the non-canonical examples of interest. We will show that the familiar leapfrog integrator can be represented as a degenerate variational integrator. 

We begin by reviewing the variational formulation of canonical Hamiltonian dynamics. Consider a one-degree-of-freedom Hamiltonian system described by a coordinate $q$, conjugate momentum $p$, and autonomous Hamiltonian $H(q, p)$. One degree-of-freedom and an autonomous Hamiltonian are assumed for simplicity in this Subsection; the generalizations are straightforward and will be encountered in later subsections. The equations of motion may be derived from an action principle employing the following Lagrangian \cite{Arnold_1989_phase_space_action, Goldstein_2001_phase_space_action}:
  \begin{equation}
    \label{eq:canonical_ps_lagrangian}
    L(q, p, \dot{q}, \dot{p}) = p \dot{q} - H(q, p).
  \end{equation}
Because this Lagrangian acts on points $(q, p)$ in the Hamiltonian phase space, it is referred to as a ``phase-space Lagrangian'' \cite{Littlejohn_1983, Cary_1983}. Equation~\ref{eq:canonical_ps_lagrangian} is intimately related to the standard Legendre transform relationship between a Hamiltonian and a Lagrangian:
\begin{equation}
  \label{eq:legendre_transform}
  L(q, \dot{q}) = p(q, \dot{q}) \dot{q} - H(q, p(q, \dot{q})),  
\end{equation}
where $p(q, \dot{q})$ is a function determined by inverting $\dot{q} = \partiald{H}{p}(q, p)$. Note that a phase-space description is retained in Eq.~\eqref{eq:canonical_ps_lagrangian} by treating $q$ and $p$ as \emph{independent} coordinates until the Euler-Lagrange equations inform us of their relationship.

To identify Euler-Lagrange equations corresponding to this Lagrangian, consider a path $(q(t), p(t))$ for $t \in [0, T]$ and define an action $S$ acting on the path $(q, p)$ according to
\begin{equation}
  \label{eq:ps_action}
  S(q, p) = \int_0^T L(q(t), p(t), \dot{q}(t), \dot{p}(t)) \d{t}.
\end{equation}
Hamilton's principle of least action states that the true trajectory extremizes the action functional $S$. Varying the action with respect to the path $(q, p)$, one obtains
\begin{align}
& \delta S(q, p) =  \exteriord S (q, p) \cdot \vectorTwo{\delta q}{\delta p} = \nonumber \\ &\quad \int_0^T\bigg[ \big(\dot{q}(t) - H_{,p}(q(t), p(t)) \big) \delta p(t) - \nonumber \\
& \quad \quad \big(\dot{p}(t) + H_{,q}(q(t), p(t)) \big) \delta q(t) \bigg] \d{t} + p \delta q \vert^{t=T}_{t=0}
\label{eq:ps_action_variation}
\end{align}
where $_{,q}$ denotes differentiation with respect to $q$, for example, and we have used integration by parts to obtain the result. Asserting that variations at the endpoints are zero\footnote{The fixed endpoint condition introduces technical nuance on the existence of a path connecting such endpoints, especially for phase-space and degenerate Lagrangians. See Ref.~\onlinecite{Tyranowski_2014} for technical details.}, the action is  extremized by trajectories obeying the following Euler-Lagrange equations:
\begin{subequations}
  \label{eq:canonical_ps_ele}
\begin{align}
  \dot{q} - H_{,p}(q, p) & = 0, \\
   -\dot{p} - H_{,q}(q, p) & = 0,
  \end{align}
\end{subequations}
for all $t$ in $[0, T]$. The canonical phase-space Lagrangian Eq.~\eqref{eq:canonical_ps_lagrangian} therefore allows one to derive Hamilton's equations (in canonical coordinates) as the Euler-Lagrange equations of an action principle formulated in phase space \cite{Goldstein_2001_phase_space_action}.

The emergence of a system of first-order ODEs --- rather than a system of second-order ODEs --- as Euler-Lagrange equations is one indication that the phase-space Lagrangian of Eq.~\eqref{eq:canonical_ps_lagrangian} is \emph{degenerate}. To address degeneracy in general, let $z$ represent the generalized coordinates of the Lagrangian, $L(z, \dot{z})$. Here, $z$ is chosen to generalize across the cases of interest for this manuscript; in the canonical setting $z=(q,p)$. A Lagrangian $L(z, \dot{z})$ is defined to be \emph{degenerate} if
\begin{equation}
  \label{eq:degenerate_l}
  \det\left( \frac{ \partial^2 L}{\partial \dot{z} \partial \dot{z}} \right) = 0.
\end{equation}
In general, degeneracy is a local property of the Lagrangian, but we will simplify the discussion by assuming the Lagrangian has a globally constant degree of degeneracy (which is true for all Lagrangians considered in this work). To understand the relationship between degeneracy and the order of the Euler-Lagrange system of equations, recall that the Euler-Lagrange equations are in general
\begin{equation}
  \partiald{L}{z} - \ddt \partiald{L}{\dot{z}} = 0,
\end{equation}
or expanding the time derivative:
\begin{equation}
  \partiald{L}{z} - \frac{\partial^2 L}{\partial \dot{z} \partial z} \cdot \dot{z} - \frac{\partial^2 L}{\partial \dot{z} \partial \dot{z} } \cdot \ddot{z} = 0.
  \label{eq:euler-lagrange}  
\end{equation}
The final term reveals that one is able to uniquely solve for $\ddot{z}(z, \dot{z})$ if and only if the Lagrangian is  non-degenerate, or \emph{regular}. If the Hessian matrix $\frac{\partial^2 L}{\partial \dot{z} \partial \dot{z} }$ is not full rank, the order of the Euler-Lagrange ODE system will be reduced. Examining the phase-space Lagrangian Eq.~\eqref{eq:canonical_ps_lagrangian}, the Hessian  $\frac{\partial^2 L}{\partial \dot{z} \partial \dot{z} }$ is completely zero, so no second-order time derivatives appear in Eq.~\eqref{eq:canonical_ps_ele} at all. Indeed, the intent of the phase-space Lagrangian is to recover a system of first-order ODEs. Contrast this with a (``configuration-space'') Lagrangian of the form:
\begin{equation}
  L(q, \dot{q}) = \half \dot{q}^2 - V(q),
\end{equation}
which is \emph{not} degenerate and yields a (single) second-order Euler-Lagrange equation. For an interesting example of a degenerate Lagrangian that is not a phase-space Lagrangian, see Fermat's principle in Appendix~\ref{sec:fermats_principle}.

\begin{figure}
  \includegraphics[width=0.35\textwidth]{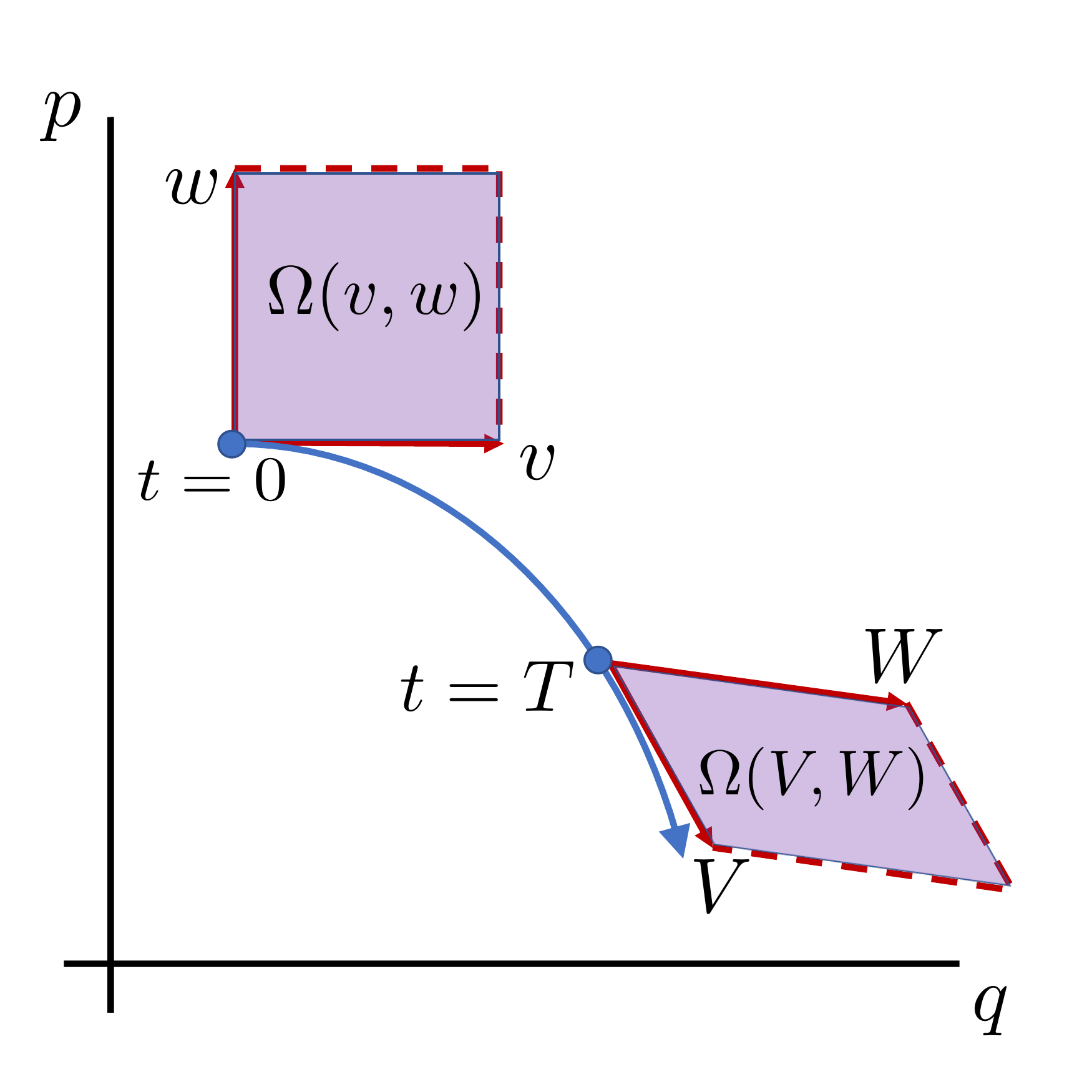}
  \caption{Canonical Hamiltonian systems preserve phase-space area. Given two vectors $(v, w)$ at some initial point at time $t=0$, the area they span is given by $\Omega(v, w)$. At a later time $t=T$, the vectors have evolved to $(V, W)$ at some other point in the $(q, p)$ plane, but the area they span $\Omega(V, W)$ remains constant. Because this is true for any $v, w$ and any time $T$, this property is referred to as preserving the symplectic structure $\Omega$. In multiple degrees of freedom, the sum of the areas in each of the $q^i, p_i$ planes is preserved.}
  \label{fig:symplectic_area_canonical}
\end{figure}

An important property of Hamiltonian systems is that they preserve a \emph{symplectic structure} \cite{Marsden_1999, Holm_2009}. In the canonical setting, this means that areas in the $(q, p)$ Hamiltonian phase-space are preserved as they are evolved according to the flow of Hamilton's equations. This property may be rapidly verified using the phase-space action principle and tools from differential geometry, which are summarized in Appendix~\ref{sec:differential_geometry_primer}. To do so, we consider the action in Eq.~\eqref{eq:ps_action_variation} \emph{restricted} to act only on trajectories satisfying the Euler-Lagrange equations, Eq.~\eqref{eq:canonical_ps_ele}. The restricted action $\bar{S}$ can then be considered to be a function of the initial conditions, $(q(0), p(0))$, with the remainder of the path being determined by the solution of the Euler-Lagrange equations. Taking an exterior derivative (see \ref{sec:differential_geometry_primer}), the integrand in Eq.~\eqref{eq:ps_action_variation} is zero, so only the boundary terms from the integration by parts remain:
\begin{equation}
  \label{eq:ds_boundary_canonical}
  \exteriord \bar{S}(q(0), p(0)) = p(T) \exteriord q(T) - p(0) \exteriord q(0),  
\end{equation}
where again $(q(T), p(T)$ are determined by the initial conditions according to the Euler-Lagrange equations. We then take a second exterior derivative combined with the property that $\exteriord^2 = 0$ (a differential geometric analog of $\nabla \times \nabla = 0$ and $\nabla \cdot \nabla \times = 0$) to show
\begin{equation}
  \label{eq:canonical_symplecticity}
  \exteriord q \wedge \exteriord p \vert_{t=0} = \exteriord q \wedge \exteriord p \vert_{t=T}.
\end{equation}
That is, the solutions of Hamilton's equations preserve the ``differential two-form'' $\Omega = \exteriord q \wedge \exteriord p$. As discussed in the Appendix, this two-form is a twice-covariant anti-symmetric tensor that may be interpreted as calculating the area spanned by two vectors in the Hamiltonian phase-space. The fact that solutions to Hamilton's equations preserve this symplectic structure means that as two arbitrary vectors are evolved along the solution to Hamilton's equations, the area they span will remain constant; see Fig.~\ref{fig:symplectic_area_canonical}. 

Turning now to the construction of numerical solutions to Eq.~\eqref{eq:canonical_ps_ele}, it is desirable to choose a numerical method that also preserves these phase-space areas, i.e., a \emph{symplectic integrator}. Symplectic integrators generate trajectories that are nearby to \emph{some} Hamiltonian system that converges to the original as the step size approaches zero (provided the method is consistent) \cite{Hairer_2006_symplectic, McLachlan_2006}. If the Hamiltonian of the continuous system is $H(q,p)$, then a backward error analysis reveals that the numerical solution after one time step is the time-$h$ solution to a Hamiltonian system of the form:
\begin{equation}
  \label{eq:modified_hamiltonian}
  \widetilde{H}(q, p) = H(q, p) + h H_1(q, p) + h^2 H_2(q, p) + ..., 
\end{equation}
where the power series in $h$ is asymptotic \cite{Hairer_2006_symplectic} and the $H_i$ functions depend on $H$ and its derivatives. Although the solution is an approximation of the true dynamics, symplectic integrators ensure that the \emph{character} of the solution remains Hamiltonian. This leads to many desirable properties, including bounded energy errors (for systems with time-independent Hamiltonians), subject to the technical qualifications of the asymptotic expansion above.

One way to systematically construct algorithms that preserve \emph{some} symplectic structure is to use a technique known as variational integration, wherein all discrete-time approximations are introduced to the Lagrangian and the action \cite{Marsden_2001}. The numerical algorithm is determined by requiring the numerical trajectory (a path that is now discrete in time) to extremize the discrete-time action. The determination of a symplectic structure preserved by the variational integrator then follows the same procedure as used to determine Eq.~\eqref{eq:canonical_symplecticity}. As an illustration of this variational integration procedure, and an example of what can go wrong in the context of degenerate Lagrangian systems, let us begin by constructing a discrete-time approximation to canonical phase-space Lagrangian in Eq.~\eqref{eq:canonical_ps_lagrangian}. Specifically, consider the following ``midpoint'' \emph{discrete Lagrangian}:
\begin{align}
&L_d(z_k, z_{k+1}) = \nonumber \\
& \half \left(L\left(z_k, \frac{z_{k+1} - z_k}{h} \right) + L\left(z_{k+1}, \frac{z_{k+1} - z_k}{h} \right) \right) = \nonumber \\
                & \frac{p_{k} + p_{k+1}}{2}\frac{q_{k+1} - q_{k}}{h} - \half \left( H(q_k, p_k) + H(q_{k+1}, p_{k+1}) \right),
\label{eq:midpoint_canonical_ld}
\end{align}
where $z = (q, p)^T$, $z_k$ denotes the numerical solution at time $t_k$ and the timestep size is $h$. The specific choice of discretization is plausible; one might choose such a time-centered discretization to obtain a time-centered (and therefore second-order accurate) algorithm. The \emph{discrete action} $S_d$ corresponding to this discrete Lagrangian is a summation over the time interval:
\begin{equation}
  \label{eq:discrete_action}
  S_d(z_0, z_1, ..., z_N) = \sum_{k=0}^{N-1} h L_d(z_k, z_{k+1}),
\end{equation}
where the time interval $[0, T]$ has been divided into $N$ increments of equal size $h$. This summation is clearly a discretized version of the integral in Eq.~\eqref{eq:ps_action}. A discrete analog of the Euler-Lagrange equations, dubbed the \emph{discrete Euler-Lagrange equations}, is obtained by requiring the discrete action to be stationary with respect to variations in $z_k$ for all $k = 1, ..., N-1$, so
\begin{align}
  \label{eq:dsd_general}
 \delta S_d = & h \sum_{k=1}^{N-1} \left(  \frac{ \partial L_d(z_{k-1}, z_{k})}{ \partial z_k} + \frac{ \partial L_d(z_{k}, z_{k+1})}{ \partial z_k} \right) \cdot \delta z_k + \nonumber \\
& \quad h \frac{\partial L_d(z_0, z_1)}{\partial z_0} \cdot \delta z_0 + h \frac{\partial L_d(z_{N-1}, z_N)}{\partial z_N} \cdot \delta z_N.
\end{align}
Again asserting the variations are zero at the endpoints, the discrete action is extremized by discrete trajectories satisfying the discrete \emph{Euler-Lagrange equations}: 
\begin{equation}
  \label{eq:discrete_euler-lagrange}
  \partiald{L_d(z_{k-1}, z_k)}{z_k} + \partiald{L_d(z_k, z_{k+1})}{z_k} = 0,
\end{equation}
for all $k=1, ..., N-1$. For the midpoint discrete Lagrangian in Eq.~\eqref{eq:midpoint_canonical_ld}, the discrete Euler-Lagrange equations are:
\begin{subequations} 
  \label{eq:canonical_explicit_midpoint}
  \begin{align}
    \quad -p_{k+1} + p_{k-1} - 2 h H_{,q}(q_k, p_k) & = 0, \\
    \quad q_{k+1} - q_{k-1} - 2 h H_{,p}(q_k, p_k) & = 0.
  \end{align}
\end{subequations}

Immediately, we observe that something went awry during the discretization procedure. Whereas the continuous equations of motion, Eq.~\eqref{eq:canonical_ps_ele}, are a system of first-order differential equations, the discrete Euler-Lagrange equations above manifest a system of \emph{second-order} difference schemes: to determine $(q_{k+1}, p_{k+1})$, one must supply both $(q_{k-1}, p_{k-1})$ and $(q_k, p_k)$. Because of this discrepancy between the order of the continuous and discrete equations, Eq.~\eqref{eq:canonical_explicit_midpoint} is referred to as a \emph{multistep method} \cite{Dahlquist_1956, Hairer_2006_multistep}; it is a two-step method for solving a system of first-order differential equations. This is not to be confused with \emph{multistage} methods, such as Runge-Kutta schemes, which evaluate the ODE vector field at multiple intermediate stages of a timestep but do not require additional initial conditions. The particular multistep scheme obtained above is referred to as the ``explicit midpoint scheme'' \cite{Hairer_2006_multistep}. 

Because the explicit midpoint scheme is a higher-order difference system than the continuous equations it models, the numerical trajectory it generates contains additional modes not present in the continuum dynamics. These additional ``parasitic'' or ``computational'' modes \cite{Hairer_2006_multistep, Hairer_1999} have an interrelated, deleterious effect on the conservation properties of the variational integrator and the stability of the numerical trajectories it generates. Beginning with the conservation properties, the motivation for deriving an integrator from a discrete variational principle is to obtain an area-preserving result analogous to that of Eq.~\eqref{eq:canonical_symplecticity} \cite{Marsden_2001}. To analyze the conservation properties of the explicit midpoint scheme~\eqref{eq:canonical_explicit_midpoint}, we again refer to the variational principle. Consider Eq.~\eqref{eq:dsd_general} for the midpoint discrete Lagrangian and restrict it to trajectories satisfying the discrete Euler Lagrange equations. Analogous to Eq.~\eqref{eq:ds_boundary_canonical}, taking a derivative of the \emph{restricted discrete action} $\bar{S}_d$ recovers only the boundary terms (i.e. the $\delta z_0, \delta z_N$ terms) in Eq.~\eqref{eq:dsd_general}:
\begin{align}
  \exteriord \bar{S}_d(q_0, p_0, q_1, p_1) & = \frac{1}{2}(q_1 - q_0) \exteriord p_0 - \frac{1}{2}(p_0 + p_1) \exteriord q_0 - \nonumber \\
& \quad \frac{h}{2} H_{,q}(q_0, p_0) \exteriord q_0 - \frac{h}{2} H_{,p}(q_0, p_0) \exteriord p_0 + \nonumber \\
  & \quad \frac{1}{2}(q_N - q_{N-1}) \exteriord p_N - \frac{1}{2}(p_{N-1} + p_N) \exteriord q_N - \nonumber \\
& \quad \frac{h}{2} H_{,q}(q_N, p_N) \exteriord q_N - \frac{h}{2} H_{,p}(q_N, p_N) \exteriord p_N,
\end{align}
where $(q_{N-1}, p_{N-1}, q_N, p_N)$ are determined by the initial condition $(q_0, p_0, q_1, p_1)$ by iterating the discrete Euler-Lagrange time advance. Taking a second exterior derivative (i.e. applying Eq.~\eqref{eq:exteriord_operation}) identifies a symplectic structure preserved by the variational integrator:
\begin{equation}
  \label{eq:multistep_omega}
  \exteriord q_1 \wedge \exteriord p_0 + \exteriord q_0 \wedge \exteriord p_1 = \exteriord q_{N-1} \wedge \exteriord p_N + \exteriord q_N \wedge \exteriord p_{N-1},
\end{equation}
where the terms involving the Hamiltonian have become zero due to antisymmetry and the equivalence of mixed partial derivatives. Although this symplectic two-form resembles that of the continuous system ($\exteriord q \wedge \exteriord p$), it resides on a space \emph{twice} as large as the original Hamiltonian phase space! The two-form in Eq.~\eqref{eq:canonical_symplecticity} and Fig.~\ref{fig:symplectic_area_canonical} is on a two-dimensional space (with coordinates $(q, p)$), whereas this two-form is on a four-dimensional space (with coordinates $(q_0, p_0, q_1, p_1)$). Because the multistep variational integrator preserves areas in a higher dimensional space, we cannot expect it to behave the same as familiar symplectic integrators; that is, we cannot expect the numerical trajectory to be a solution to a Hamiltonian system of the form of Eq.~\eqref{eq:modified_hamiltonian}.

Although the multistep variational integrator does not preserve the correct areas, it remains possible at this stage of reasoning that the numerical method behaves quite well. The presence of parasitic modes does not imply instability; there exist entire families of multistep schemes for which the modes are entirely well behaved (i.e., all parasitic modes are damped) \cite{Bashforth_1883, Dahlquist_1956, Hairer_1999}. Optimistically, one may hope that the rather unusual symplectic structure in Eq.~\eqref{eq:multistep_omega} serves to restrict the behavior of the undesired modes. Unfortunately, such optimism is rapidly dissuaded by a numerical example.

\begin{figure}	\includegraphics[width=0.4\textwidth]{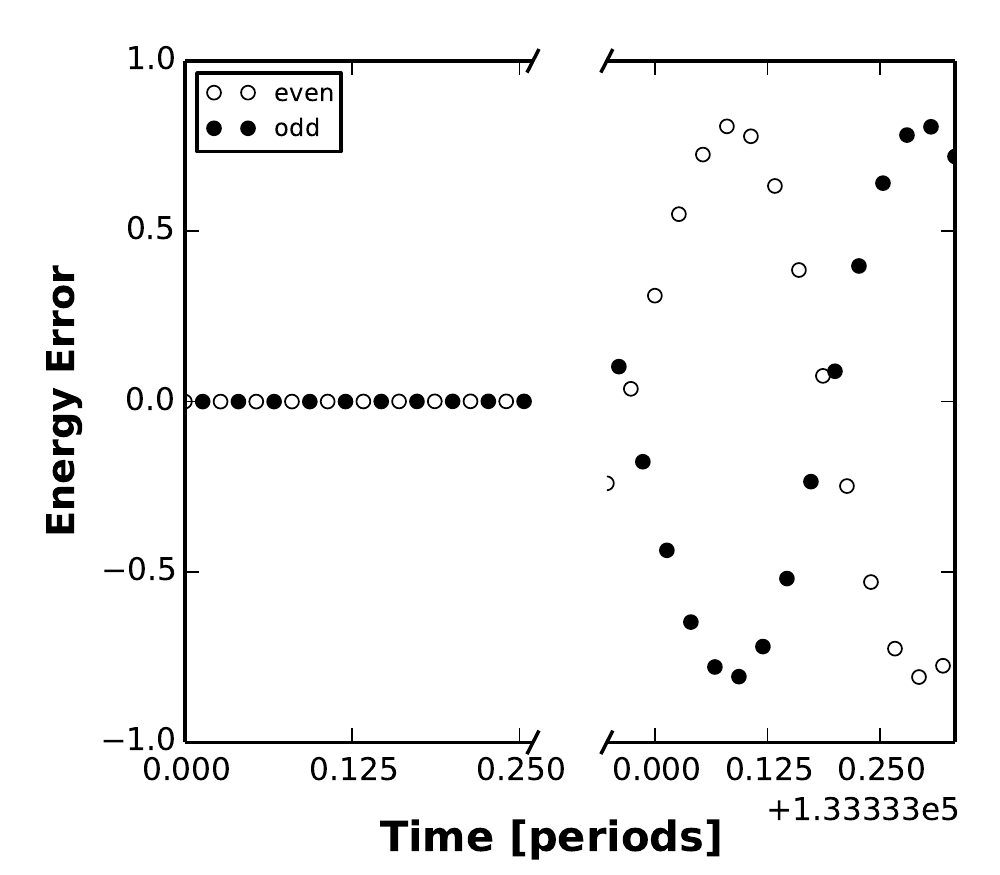}
	\caption{The two-step variational integrator, Eq.~\eqref{eq:canonical_explicit_midpoint}, admits parasitic mode instabilities when applied to the nonlinear pendulum problem. At early times, the even- and odd-indexed points in the trajectory lie on a smooth curve. After $\approx 10^5$ oscillation periods, a large even-odd oscillation distorts the trajectory, as evidenced by the order-unity energy error. Initial condition: $(q, p) = (1, 0)$; timestep $h = 0.1$.}
	\label{fig:parasitic_pendulum}
\end{figure}

Figure~\ref{fig:parasitic_pendulum} depicts the nonlinear pendulum system, $H(q, p) = p^2/2 + 1 - \cos(q)$, integrated using the two-step variational integrator, explicit midpoint, Eq.~\eqref{eq:canonical_explicit_midpoint}. To highlight the presence of the spurious numerical mode, the even- and odd-numbered steps are distinguished with white and black markers, respectively. Of course, the distinction between even- and odd-indexed times is purely a feature of the time discretization and cannot pertain to physical reality. At early times, the trajectory appears to be smooth and a good representation of the physical dynamics. As time progresses, however, the presence of unphysical modes becomes apparent as the even- and odd-indexed trajectories diverge. The even-odd character arises because the parasitic modes correspond to eigenvalues near negative one, in a linear stability analysis. If the modes are linearly or nonlinearly unstable, the modes can grow to large amplitude. The presence of such a numerical instability in Fig.~\ref{fig:parasitic_pendulum} indicates that the four-dimensional symplectic structure in Eq.~\eqref{eq:multistep_omega} is insufficient for obtaining the desired long-term numerical fidelity. Moreover, such instability in multistep variational integrators is not limited to this particular example. It has been recently shown that \emph{any} variationally-derived multistep method cannot have parasitic modes that are all damped \cite{Ellison_thesis}; if one parasitic mode is damped, there exists another parasitic mode that is unstable and amplified in time. The best prospect is to eliminate the parasitic modes altogether.

Interestingly, it is not especially difficult to select a discretization of the phase-space Lagrangian that eliminates the parasitic modes in the canonical setting. Instead of choosing the midpoint discretization of Eq.~\eqref{eq:midpoint_canonical_ld}, consider the following discretization:
\begin{equation}
  \label{eq:canonical_forward_ld}
  L_d(q_k, p_k, q_{k+1}, p_{k+1}) = p_k \frac{q_{k+1} - q_k}{h} - H(q_{k+1}, p_k).
\end{equation}
Proceeding to vary the discrete action $S_d$ of Eq.~\eqref{eq:discrete_action}
\small
\begin{align}
& \delta S_d(q_0, p_0, q_1, p_1, ..., q_N, p_N) = \nonumber \\
& \sum_{k=0}^{N-1} \left(q_{k+1} - q_k - h H_{,p}(q_{k+1}, p_k) \right) \delta p_k + \nonumber \\
& \quad \left(p_k - h H_{,q}(q_{k+1}, p_k) \right) \delta q_{k+1} - p_k \delta q_k = \nonumber \\
& \sum_{k=0}^{N-1}  \bigg[ \left(q_{k+1} - q_k - h H_{,p}(q_{k+1}, p_k) \right) \delta p_k + \nonumber \\
& \quad \left(p_k - p_{k+1} - h H_{,q}(q_{k+1}, p_k) \right) \delta q_{k+1} \bigg] - p_0 \delta q_0 + p_N \delta q_N.
\end{align}
\normalsize
The discrete Euler-Lagrange equations for this system are then
\begin{subequations}
  \label{eq:forward_canonical_del}
  \begin{align}
   \quad p_{k} - p_{k+1} - h H_{,q}(q_{k+1}, p_{k}) & = 0, \label{eq:se_top} \\
   \quad q_{k+1} - q_k - h H_{,p}(q_{k+1}, p_k) & = 0. \label{eq:se_bottom}
  \end{align}
\end{subequations}
In contrast to the first variational integrator, this algorithm is a one-step method, requiring only a single initial condition $(q_0, p_0)$. No parasitic modes can possibly be present because the order of the difference equations matches the order of the ODE system. This variational integrator is also well known; it is the first-order-accurate symplectic Euler scheme \cite{Hairer_2006_seuler}. It may be solved by first applying Eq.~\eqref{eq:se_bottom} to implicitly determine $q_{k+1}$, then applying Eq.~\eqref{eq:se_top} to explicitly identify $p_{k+1}$. For separable Hamiltonian systems,  i.e., Hamiltonians of the form $H(q, p) = K(p) + V(q)$, this is the same as a leapfrog scheme if one instead interprets the momentum coordinates as ``staggered'': $p_{k} \mapsto p_{k+1/2}$. This variational integrator may be shown to be symplectic by again appealing to the variational principle. Restricting attention to trajectories that satisfy the discrete Euler-Lagrange equations~\eqref{eq:forward_canonical_del}, the derivative of the restricted discrete action leaves only the boundary terms:
\begin{equation}
  \label{eq:dsd_canonical}
  \exteriord \bar{S}_d(q_0, p_0) = -p_0 \exteriord q_0 + p_N \exteriord q_N.
\end{equation}
Taking a second exterior derivative recovers the desired result:
\begin{equation}
  \exteriord q_0 \wedge \exteriord p_0 = \exteriord q_N \wedge \exteriord p_N.  
\end{equation}
This variational integrator therefore preserves the \emph{same} symplectic two-form as the true Hamiltonian system; both the symplectic Euler scheme and the leapfrog advance are well known to be symplectic.

Although the parasitic modes have been eliminated, the preceding variational integrator is not centered in time and is only first-order accurate. It is of course desirable to achieve second-order accuracy, which could be approached in a number of ways. Choosing a time-centered discretization apparently introduces parasitic mode instabilities, as shown with the explicit midpoint method. An alternative route to time symmetrization is to alternate the time advance between a first-order scheme and its \emph{adjoint}. The adjoint scheme results from transforming $h \mapsto -h$ and swapping $k$ and $k+1$ in the discrete Lagrangian\cite{Marsden_2001}: 
\begin{equation}
  \label{eq:canonical_backward_ld}
L_d(q_k, p_k, q_{k+1}, p_{k+1}) = p_{k+1} \frac{q_{k+1} - q_k}{h} - H(q_{k}, p_{k+1}).
\end{equation}
The resulting variational integrator will be the other symplectic Euler scheme --- the adjoint of the previous method, involving implicit determination of $p_{k+1}$ followed by explicit identification of $q_{k+1}$ \cite{Hairer_2006_adjoint}. By alternating between the two symplectic Euler schemes, the composition is time symmetric and therefore second-order accurate. For this system, both of the symplectic Euler integrators used in the composition preserve the same (canonical) symplectic structure, so their composition also preserves the canonical symplectic structure and therefore exhibits the expected long-term numerical fidelity. Later we will encounter degenerate variational integrators that preserve \emph{different} symplectic structures than their adjoint, so it will not be as clear what can be rigorously claimed about the conservation properties of such accuracy-enhancing compositions.

The most important lesson from these examples is that simply changing the choice of discrete Lagrangian recovered a system of first-order difference schemes, thereby eliminating the parasitic mode instabilities and preserving a symplectic two-form on the Hamiltonian phase space. The question inspired by these studies becomes: \emph{which} discrete Lagrangians will yield one-step methods? This question has been addressed in detail in a recent doctoral thesis \cite{Ellison_thesis}, and in brief may be explained as a result of degeneracy in the \emph{discrete} Lagrangian. To examine degeneracy in the discrete setting, let us return to generalized coordinates $z$ representing the arguments of the Lagrangian. A discrete Lagrangian typically depends on $z_k$ and $z_{k+1}$, so $L_d = L_d(z_k, z_{k+1})$. The discrete Lagrangian is defined to be \emph{degenerate} if \cite{Marsden_2001}
\begin{equation}
  \label{eq:discrete_degeneracy}
  \det \left( \frac{\partial^2 L_d}{\partial z_k \partial z_{k+1}} \right) = 0.
\end{equation}
Just like the continuous degeneracy condition, the degeneracy condition for a discrete Lagrangian may be understood as a solvability condition for the discrete Euler-Lagrange equations, Eq.~\eqref{eq:discrete_euler-lagrange}. According to the implicit function theorem, a necessary and sufficient condition for solving for $z_{k+1}$ as a function of $(z_{k}, z_{k-1})$ is that the discrete Lagrangian is non-degenerate. That is, the discrete Euler-Lagrange equations are a well-defined system of second-order difference schemes if and only if the discrete Lagrangian is non-degenerate.

A cursory inspection of Eq.~\eqref{eq:discrete_euler-lagrange} might lead one to conclude that the discrete Euler-Lagrange equations do not specify any time advance if the discrete Lagrangian is degenerate. However, the experiment yielding the symplectic Euler/leapfrog scheme proves that --- much like the phase-space Lagrangian --- degeneracy can correspond to a \emph{reduction} in the \emph{order} of the system of difference equations. In fact, the discrete degeneracy condition in Eq.~\eqref{eq:discrete_degeneracy} serves as a useful guide for discerning which Lagrangians will yield a multistep method from those that yield a reduced order system of difference equations. For example, the midpoint discrete Lagrangian Eq.~\eqref{eq:midpoint_canonical_ld} is \emph{not} degenerate:
\begin{equation}
  \frac{ \partial^2 L_d(z_k, z_{k+1})}{\partial z_k \partial z_{k+1}} = \matrixTwobyTwo{0}{\frac{-1}{2h}}{\frac{1}{2h}}{0},
\end{equation}
ensuring a system of second-order difference equations (Eq.~\eqref{eq:canonical_explicit_midpoint}), which exceeded the order of the original differential equations (Eq.~\eqref{eq:canonical_ps_ele}). In contrast, the discrete Lagrangian in Eq.~\eqref{eq:canonical_forward_ld}, which yielded the symplectic Euler/leapfrog advance, \emph{is} degenerate:
\begin{equation}
  \frac{ \partial^2 L_d(z_k, z_{k+1})}{\partial z_k \partial z_{k+1}} = \matrixTwobyTwo{0}{0}{\frac{1}{h} - H_{,pq}(q_{k+1}, p_{k})}{0},
\end{equation}
so the rank is one and the determinant is zero. This degeneracy indicates that the discrete Euler-Lagrange equations cannot be a system of second-order difference equations; instead, we identified a one-step method. 

In general, matching the order of the difference equations to the order of the differential equations requires the two systems to be degenerate to ``the same degree''. It is not sufficient to simply ensure that both the continuous and discrete Lagrangians are degenerate, but rather one must ensure the orders of the relevant systems of equations are the same. Indeed, one could imagine a phase-space Lagrangian for a two-degree-of-freedom Hamiltonian system in which the degrees of freedom are uncoupled. One could then choose a discrete Lagrangian that is degenerate in one of the degrees of freedom and non-degenerate in the other. Although the overall discrete Lagrangian would be degenerate, it would not be degenerate \emph{enough} to eliminate \emph{all} of the parasitic modes. Let us then call a discrete Lagrangian that is degenerate and whose discrete Euler-Lagrange equations have the same order as the continuous system \emph{properly degenerate}.

The condition for an arbitrary discrete Lagrangian to be properly degenerate is discussed in general in a recent thesis\cite{Ellison_thesis}. For the scope of this manuscript, wherein we are only interested in discretizing phase-space Lagrangians, we claim without proof that a discrete Lagrangian is properly degenerate if the rank of the Hessian tensor in Eq.~\eqref{eq:discrete_degeneracy} equals the number of degrees of freedom. This condition will serve as a guide for detecting that the desired reduction in the order of the numerical system has taken place. In lieu of proving this order matching condition, we will show case-by-case that the presented integrators satisfy the condition, are one-step methods, and preserve symplectic structures on the Hamiltonian phase space. Note that it is not yet known how to systematically construct variational integrators satisfying the proper degeneracy condition, but the condition remains useful nonetheless for rapidly assessing whether a chosen discrete Lagrangian will yield a one-step method.

The ensuing subsections build upon the intuition established above, striving to develop variational integrators with the proper degree of degeneracy for the important applications of magnetic field line flow and guiding center trajectories. These two Hamiltonian systems also stem from phase-space Lagrangians, albeit in forms more general than the canonical phase-space Lagrangian. To see how these more general \emph{non-canonical} phase-space Lagrangians might come about, consider an arbitrary coordinate transformation of the form
\begin{equation*}
  (q, p) \mapsto z(q, p).
\end{equation*}
In this case, the phase-space Lagrangian becomes
\begin{align}
  L(z, \dot{z}) & = p(z) \cdot \partiald{q}{z} \cdot \dot{z} - H(z) \nonumber \\
                & = \vartheta(z) \cdot \dot{z} - H(z).
  \label{eq:noncanonical_ps_lagrangian}  
\end{align}
Both the magnetic field line and guiding center Lagrangians are in the form of Eq.~\eqref{eq:noncanonical_ps_lagrangian}. Indeed, such a ``non-canonical coordinate transformation'' illustrated above plays a central role in the derivation of Hamiltonian guiding center dynamics \cite{Littlejohn_1983, Cary_2009}. For non-canonical phase-space Lagrangians, achieving a one-step (degenerate) variational integrator does not immediately follow from such a simple discretization as the one that yielded the symplectic-Euler integrator. However, the following sections will employ a combination of electromagnetic gauge transformations and assumptions on the coordinates to facilitate the desired result of one-step variational integrators in the non-canonical phase-space coordinates.

\subsection{Magnetic Field Line DVI}
\label{ssec:magnetic_field_line_flow}

Consider a time-independent magnetic vector potential $A(x)$ and a corresponding magnetic field $B = \nabla \times A$. (We may be interested in time-dependent fields, but would only consider a single instant in time when tracing field lines.) One can trace or ``follow'' magnetic field lines by solving the differential equation:
\begin{equation}
  \label{eq:standard_field_line_ode}
  \totald{x}{\tau} = B(x),
\end{equation}
where $\tau$ parameterizes the distance along the field line from some initial condition $x_0$. It is possible to obtain these equations - up to some path parameterization choices - from action principles in phase space \cite{Cary_1983}. As a starting point, consider the Lagrangian:
\begin{equation}
  \label{eq:vdota_lagrangian}
  L(x, \dot{x}) = A(x) \cdot \dot{x},
\end{equation}
where the dot now denotes the derivative with respect to the path length variable $\tau$. The Euler-Lagrange equations corresponding to this Lagrangian are given by
\begin{equation}
\label{eq:ele_vdota_lagrangian}
\dot{x} \times B = 0.
\end{equation}

These equations inform us that we cannot move perpendicular to magnetic field lines, but the rate of traversal along the field line is left ambiguous. This ambiguity is due to the fact that the action formed from the Lagrangian in Eq.~\eqref{eq:vdota_lagrangian} admits arbitrary reparameterizations of the time coordinate:
\begin{equation*}
  \int A(x) \cdot \totald{x}{\tau} \d{\tau} = \int A(x) \cdot \totald{x}{\tau'} \d{\tau'} = \int A(x) \cdot \d{x},
\end{equation*}
for any $\tau'(\tau)$; the final expression makes it especially clear that the action does not depend on the parameterization of the path.

\begin{figure}
  \includegraphics[width=0.4\textwidth]{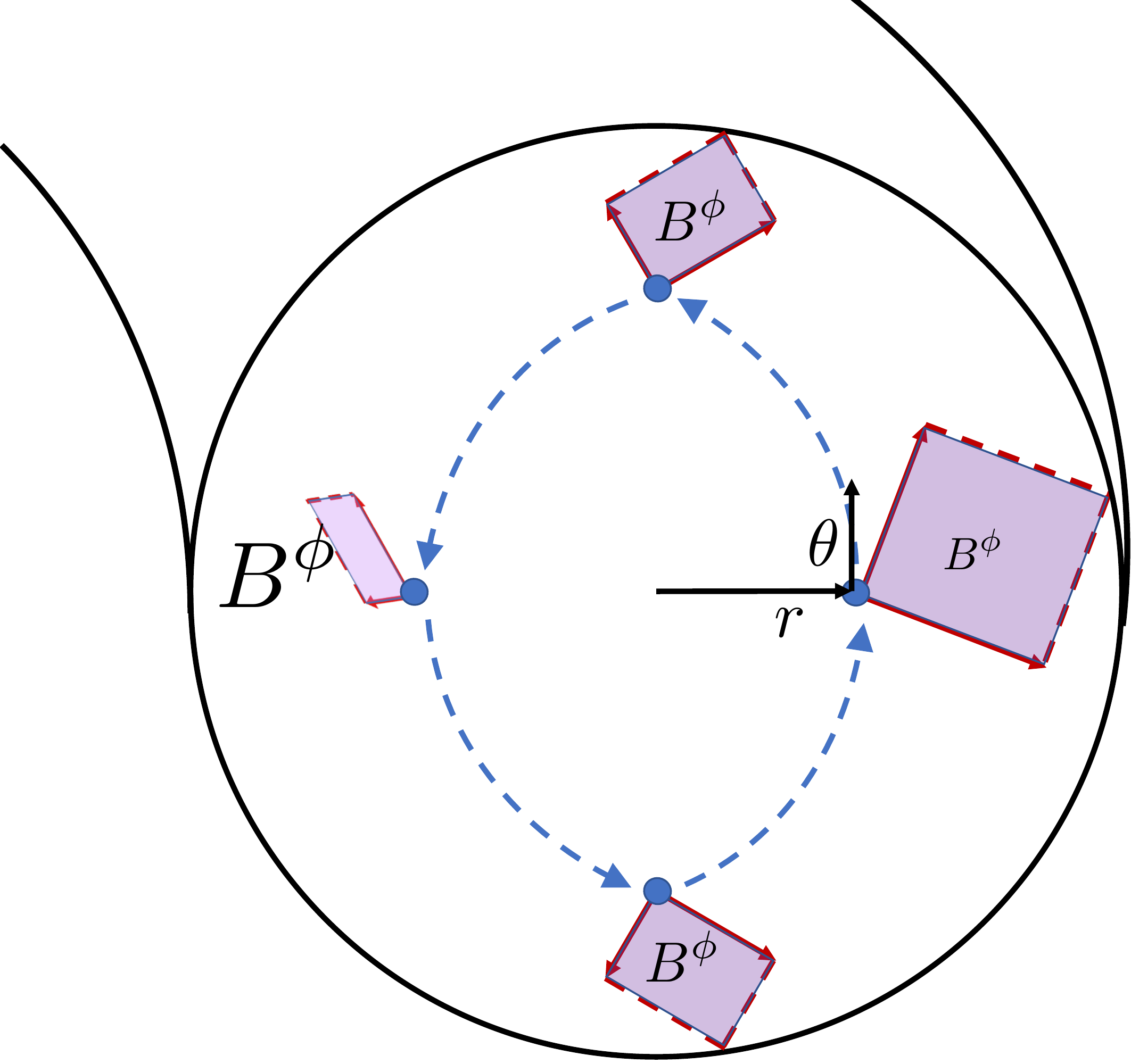}
  \caption{Magnetic field line flow preserves magnetic flux. This property is illustrated above for a tokamak with toroidal coordinates $(x^1, x^2, x^3) = (r, \theta, \phi)$. The area spanned by two vectors (shown as solid red arrows) will decrease as they move to the high-field side of a tokamak, but the toroidal flux they enclose will remain constant. }
  \label{fig:symplectic_area_bfield}
\end{figure}

To resolve this parameterization ambiguity and to reveal the underlying Hamiltonian character of this problem, \emph{choose} one of the spatial coordinates --- say $x^3$ --- to be the independent parameter. That is, we seek determination of the field line trajectory as given by the functions $x^1(x^3), x^2(x^3)$. This parameterization will remain valid as long as the contravariant $x^3$-component of $B$ is not zero. Whenever this component of the magnetic field becomes zero, one may choose a new coordinate as the independent variable to ``switch coordinate patches''. Patching issues aside, the reparameterized Lagrangian becomes
\begin{equation}
  \label{eq:mfl_lagrangian}
  L(x^1, x^2, \dot{x}^1, \dot{x}^2, x^3) = A_1 \dot{x}^1 + A_2 \dot{x}^2 + A_3,
\end{equation}
where the ``dot'' now denotes the derivative with respect to $x^3$. Now that the Lagrangian has two dynamical variables (rather than the three found in Eq.~\eqref{eq:vdota_lagrangian}), we may hope to find a Hamiltonian structure in the resulting equations. Indeed, this ``magnetic field line Lagrangian'' may be recognized as a one-degree-of-freedom phase-space Lagrangian in non-canonical coordinates, i.e. a Lagrangian in the form of Eq.~\eqref{eq:noncanonical_ps_lagrangian}, where $z = (x^1, x^2)$, $t = x^3$, $\vartheta = (A_1, A_2)$, and $H(z, t) = -A_3(x^1, x^2, x^3)$. The Euler-Lagrange equations for this phase-space Lagrangian give the non-canonical Hamilton's equations \cite{Cary_1983}:
\begin{subequations}
  \label{eq:mfl_ele}
  \begin{align}
    \dot{x}^1 & = \frac{B^1}{B^3}, \\
    \dot{x}^2 & = \frac{B^2}{B^3}.
  \end{align}
\end{subequations}
Like all phase-space Lagrangians, which yield a system of first-order differential equations, the magnetic field line Lagrangian Eq.~\eqref{eq:mfl_lagrangian} is degenerate as defined in Eq.~\eqref{eq:degenerate_l}. The symplectic structure preserved by this non-canonical Hamiltonian system is given by
\begin{equation}
  \label{eq:mfl_omega}
  \Omega = \left(A_{1,2}(x) - A_{2,1}(x) \right) \exteriord x^1 \wedge \exteriord x^2. 
\end{equation}
The conservation of this symplectic structure (and therefore flux preservation) may be derived using the phase-space variational principle in a manner exactly analogous to the derivation of Eq.~\eqref{eq:canonical_symplecticity}. As illustrated in Fig.~\ref{fig:symplectic_area_bfield}, the physical interpretation of this property is that the $B^3$ magnetic flux will be preserved as areas are evolved along with the flow of the Hamiltonian system. Whereas canonical Hamiltonian systems preserve area in the $(q, p)$-plane, magnetic field line flow preserves the $B^3$ flux in the $(x^1, x^2)$-plane. 

To construct a one-step variational integrator for this system, a properly degenerate discretization needs to be chosen for the magnetic field line Lagrangian Eq.~\eqref{eq:mfl_lagrangian}. Building upon the intuition developed in the canonical section, a reasonable first guess for a discretization that might yield a one-step method would be
\begin{align}
  \label{eq:mfl_backward_ld}
  L_d(x_{k}, x_{k+1}) =  & A_1(x_{k+1}) \frac{x^1_{k+1} - x^1_{k}}{h} + \nonumber \\
& A_2(x_{k+1}) \frac{x^2_{k+1} - x^2_{k}}{h}  + A_3(x_{k+1}).
\end{align}
However, this does not yield a one-step method or a degenerate variational integrator for a general magnetic vector potential $A$. Checking the degeneracy condition for a discrete Lagrangian (see Eq.~\eqref{eq:discrete_degeneracy}),
\begin{equation}
  \label{eq:bfield_nondegeneracy}
  \frac{ \partial^2 L_d}{\partial z_k \partial z_{k+1}} = \frac{-1}{h} \matrixTwobyTwo{A_{1,1}(x_{k+1})}{A_{1,2}(x_{k+1})}{A_{2,1}(x_{k+1})}{A_{2,2}(x_{k+1})},
\end{equation}
which is a full-rank, non-degenerate tensor for general $A$. If the discrete Lagrangian is not degenerate, then the variational integrator must be a two-step method and parasitic modes will be present; c.f. Section~\ref{ssec:canonical_hamiltonian_systems} and the numerical demonstration in Section~\ref{sec:numerical_demonstrations}. It is a straightforward exercise to calculate the discrete Euler-Lagrange equations for this discrete Lagrangian and indeed find a system of two second-order difference equations.

The form of Eq.~\eqref{eq:bfield_nondegeneracy}, however, motivates the introduction of the desired degeneracy using an electromagnetic gauge transformation. If we choose an electromagnetic gauge such that, e.g.,
\begin{equation}
  \label{eq:mfl_gauge_assumption}
  A_1 = 0,
\end{equation}
then the discrete Lagrangian in Eq.~\eqref{eq:mfl_backward_ld} reduces to
\begin{equation}
  \label{eq:ld_mfl_degenerate}
  L_d(x_k, x_{k+1}) = A_2(x_{k+1}) \frac{x_{k+1}^2 - x_k^2}{h} + A_3(x_{k+1}),
\end{equation}
which \emph{is} degenerate:
\begin{equation}
  \frac{ \partial^2 L_d}{\partial z_k \partial z_{k+1}} = \frac{-1}{h} \matrixTwobyTwo{0}{0}{A_{2,1}(x_{k+1})}{A_{2,2}(x_{k+1})}. 
\end{equation}
This is a degenerate, rank-one tensor; as claimed in Sec.~\ref{ssec:canonical_hamiltonian_systems}, when the rank of this tensor matches the number of degrees of freedom, the discrete Lagrangian is properly degenerate and the variational integrator will be a one-step method. To find such a one-step DVI, begin by varying the discrete action:
\small
\begin{align}
& \delta S_d = \nonumber \\
& \sum_{k=1}^{N-1} \bigg[ \left( A_{2,1}(x_{k}) (x_{k}^2 - x_{k-1}^2) + h A_{3,1}(x_{k}) \right) \delta x_{k}^1 + \nonumber \\
& \quad \big(A_{2,2}(x_{k}) (x_{k}^2 - x_{k-1}^2) - A_{2}(x_{k+1}) + A_2(x_k) + \nonumber \\
& \quad h A_{3,2}(x_{k}) \big) \delta x_{k}^2 \bigg] - A_2(x_{1}) \delta x_0^2 + A_2(x_{N+1}) \delta x_N^2.
\label{eq:mfl_ds}
\end{align}
\normalsize
With zero variations at the endpoints, the discrete action is extremized by the following discrete Euler-Lagrange equations:
\begin{subequations}
  \label{eq:dvi_mfl_untouched}
  \begin{align}
    A_{2,1}(x_k)\left(x_k^2 - x^2_{k-1} \right) +  h A_{3,1}(x_k)  & = 0, \label{eq:dvi_mfl_untouched_top} \\
    A_{2,2}(x_k)\left(x_k^2 - x_{k-1}^2 \right)  + A_{2}(x_{k}) - A_2(&x_{k+1}) + \nonumber \\
 h A_{3,2}(x_k) &= 0 \label{eq:dvi_mfl_untouched_bottom},
  \end{align}
\end{subequations}
for $k=1, ..., N$. Here we notice an interesting distinction from the symplectic Euler discrete Euler-Lagrange equations in  Eq.~\eqref{eq:forward_canonical_del}: the first of these equations contains evaluations at time $t_{k-1}$, $t_k$, \emph{and} $t_{k+1}$, giving the appearance of a multistep method. Guided by the degeneracy condition, however, we are confident that a reduction in order has taken place. In fact, one can formulate a one-step method from the above equations as follows: (i) Use Eq.~\eqref{eq:dvi_mfl_untouched_top} to \emph{determine} $x_{k-1}^2$ as a function of $x_k$ (ii) Replace $x^2_{k-1}$ in Eq.~\eqref{eq:dvi_mfl_untouched_bottom} with this relation and (iii) write Eq.~\eqref{eq:dvi_mfl_untouched_top} at one time index later. The result is
\begin{subequations}
\label{eq:mfl_dvi}
  \begin{align}
    A_{2,1}(x_{k+1})\left(x_{k+1}^2 - x^2_{k} \right) +  h A_{3,1}(x_{k+1})  & = 0,  \\
    - h A_{2,2}(x_{k}) \left( \frac{A_{3,1}(x_k)}{A_{2,1}(x_k)} \right) + A_2(x_k) - A_2(&x_{k+1}) + \nonumber \\
 h A_{3,2}(x_{k+1}) & = 0.
  \end{align}
\end{subequations}
That is, we have obtained a one-step variational integrator for magnetic field line flow in the $A_1 = 0$ gauge. This update is valid provided $A_{2,1} \neq 0$, which implies $B^3$ is non-zero, as previously assumed. Note that this one-step formulation generates trajectories that satisfy Eq.~\eqref{eq:dvi_mfl_untouched}; we have merely re-arranged them into a one-step method. 

It is interesting to note that one cannot express Eq.~\eqref{eq:mfl_dvi} as a direct differencing of the equations of motion Eq.~\eqref{eq:mfl_ele}. In particular, the variational integrator requires evaluations of the magnetic vector potential $A$, for example in the term $A_2(x_k)$, meaning the integrator is not gauge invariant (\emph{even} among gauges satisfying $A_1=0$). By Taylor expanding $A_2(x_{k+1}) - A_{2}(x_k)$, the gauge-dependent terms appear in the $\order{h^2}$ truncation error for this first-order method, and therefore any gauge-dependent effects will diminish as the numerical step size approaches zero. Although it is not necessarily desirable to require specification of a magnetic vector potential when numerically tracing field lines, such a gauge dependence is common among variational integrators for field-particle systems \cite{Qin_2008, Qin_2009, Squire_2012, Squire_2012_PIC}.

The symplectic structure preserved by this variational integrator may be calculated from the boundary terms in the discrete variational principle, i.e., the $\exteriord x_0^2$ and $\exteriord x_N^2$ terms in Eq.~\eqref{eq:mfl_ds}. Restricting attention to trajectories satisfying the discrete Euler-Lagrange equations, Eq.~\eqref{eq:mfl_dvi}, the derivative of the restricted action is
\begin{equation}
  \label{eq:dsd_mfl}
  \exteriord \bar{S}_d(x_0) = A_2(x_{N+1}) \exteriord x_N^2 - A_2(x_1) \exteriord x_0^2,
\end{equation}
where we recall that $x_1$ is a function of $x_0$ according to the one-step discrete Euler-Lagrange equations, Eq.~\eqref{eq:mfl_dvi}, so $x_1 = F(x_0)$ and $x_{N+1}$ actually denotes $F^{N+1}(x_0)$. Taking a second exterior derivative and $d^2 S_d = 0$ then obtains:
\begin{equation}
\label{eq:omega_d_mfl_pep}
\exteriord(A_2(F(x_0)) \exteriord x_0^2) = \exteriord(A_2(F(x_N)) \exteriord x_N^2),
\end{equation}
so $F$ preserves a symplectic structure $\Omega_d$ given by:
\begin{equation}
\label{eq:omega_d_mfl}
  \Omega_d = -\left(A_{2,1}(F(x)) \partiald{F^1}{x^1} + A_{2,2}(F(x)) \partiald{F^2}{x^1} \right) \exteriord x^1 \wedge \exteriord x^2.
\end{equation}
Comparing with Eq.~\eqref{eq:mfl_omega}, we see that the discrete symplectic structure $\Omega_d$ is not exactly the same as the continuous symplectic structure $\Omega$. They do agree, however, in the $h \rightarrow 0$ limit: as the numerical step size $h$ tends to zero, $F$ becomes the identity map, so $\partiald{F^2}{x^1} = 0$ and $\partiald{F^1}{x^1} = 1$, and of course $A_1 = 0$. In this sense, the discrete symplectic structure $\Omega_d$ may be considered to be nearby $\Omega$ for small $h$. Whereas the continuous magnetic field line flow preserves magnetic flux in the $(x^1, x^2)$-plane, the degenerate variational integrator preserves something closely related to the magnetic flux in the same $(x^1, x^2)$-plane.

The degenerate variational integrator in Eq.~\eqref{eq:mfl_dvi} is first-order accurate; in order to recover a degenerate discrete Lagrangian, we did not choose a time-symmetric discretization. After obtaining the one-step method, it is desirable to increase its order of accuracy to second-order. One approach would be to compose the DVI with its adjoint scheme, which can be derived by interchanging $k$ and $k+1$ in the discrete Lagrangian and mapping $h \mapsto -h$. The adjoint discrete Lagrangian to Eq.~\eqref{eq:ld_mfl_degenerate} is then
\begin{equation}
  \label{eq:mfl_forward_ld}
  L_d(x_k, x_{k+1}) = A_2(x_k) \frac{x_{k+1}^2 - x_k^2}{h} + A_3(x_{k}),
\end{equation}
where the $A_1=0$ gauge is still assumed. The corresponding discrete Euler-Lagrange equations are the adjoint (c.f. the discussion surrounding Eq.~\eqref{eq:canonical_backward_ld}) of the previous magnetic field DVI, Eq.~\eqref{eq:mfl_dvi}:
\begin{subequations}
\label{eq:mfl_dvi_adjoint}
  \begin{align}
    A_{2,1}(x_{k})\left(x_{k+1}^2 - x^2_{k} \right) +  h A_{3,1}(x_{k})  & = 0,  \\
    - h A_{2,2}(x_{k+1}) \left( \frac{A_{3,1}(x_{k+1})}{A_{2,1}(x_{k+1})} \right) + A_2(x_k) - &A_2(x_{k+1}) + \nonumber \\
 h A_{3,2}(x_{k}) & = 0.
  \end{align}
\end{subequations}
Second-order accuracy can certainly be achieved by composing the one-step integrator in Eq.~\eqref{eq:mfl_dvi} with the adjoint integrator in Eq.~\eqref{eq:mfl_dvi_adjoint}. If we label the first variational integrator using the map $F$ and the latter with $F^\dagger$, then a second-order accurate method will be given by $F^\dagger \circ F$. An important question, however, is what symplectic structure, is preserved by the new, time-symmetric scheme.

Typically, symplectic integrators preserve the \emph{same} symplectic structure as that of the continuous dynamics. In that case, the integrator and its adjoint both preserve the same symplectic structure and their composition is also symplectic, preserving the symplectic structure of the continuous system. Here, the symplectic structure preserved by $F^\dagger$ is
\begin{equation}
  \label{eq:omega_d_adjoint}
 \left(A_{2,2}(F^\dagger(x)) \partiald{(F^\dagger)^2}{x^1} - A_{2,1}(F^\dagger(x)) \partiald{(F^\dagger)^1}{x^1} \right) \exteriord x^1 \wedge \exteriord x^2.
\end{equation}
i.e., it is \emph{not} the same as the symplectic structure preserved by $F$, but also approaches $\Omega$ as $h$ tends to zero. It is then unclear what symplectic structure, if any, might be preserved by the composition of the two maps. To ensure the conservation properties remain intact, it may be preferable to apply an accuracy-enhancing processing technique \cite{Blanes_2004, He_2016} instead of the aforementioned composition scheme. 
\subsection{Guiding Center DVI}
\label{ssec:guiding_center_trajectories}

The final application for degenerate variational integrators is the ubiquitous guiding center system. The procedure for constructing a guiding center DVI will parallel that of the preceding sections, although a stronger restriction on the coordinates will be imposed than the electromagnetic gauge transformation employed in Section~\ref{ssec:magnetic_field_line_flow}. Specifically, it will be assumed that the magnetic vector potential $A$ and the magnetic field unit vector $b$ share a common covariant component that is zero. In the formulation presented in this paper, this condition therefore places restrictions on the form of the magnetic field as represented by the chosen coordinates. In another publication \cite{Burby_2017}, the assumption on the magnetic field is relaxed, requiring only that the contravariant toroidal component of the field be non-vanishing. Ref.~\onlinecite{Burby_2017} applies the DVI technique presented here to a re-formulated guiding center Lagrangian in which the $u b$ term is replaced by a term proportional only to $\nabla \phi$, where $\phi$ is the toroidal angle. For the scope of this contribution, we emphasize the DVI technique and restrict our attention to magnetic fields satisfying the $A_i = b_i = 0$ for some $i$. 

The variational formulation of non-canonically Hamiltonian guiding center dynamics was famously given by Littlejohn \cite{Littlejohn_1983}, who determined the following Lagrangian:
\begin{equation}
  \label{eq:gc_lagrangian}
  L(x, u, \dot{x}, \dot{u}) = \left(A(x) + u b(x) \right) \cdot \dot{x} - H_{gc}(x, u),
\end{equation}
where $u = \dot{x} \cdot b$ is the parallel velocity, $b$ is the magnetic field unit vector, and $H_{gc}$ is the guiding center Hamiltonian,
\begin{equation}
  \label{eq:gc_hamiltonian}
  H_{gc}(x, u) = \half u^2 + \mu \norm{B}(x) + \phi(x),
\end{equation}
where $\mu$ is the (constant) magnetic moment of the particle and $\norm{B}$ is the magnitude of the magnetic field. In both of these definitions, the fields have been assumed to be time independent for simplicity. Also, the vector potential $A$ has been normalized by $\frac{e}{mc}$ and the electrostatic potential $\phi$ has been normalized by $\frac{e}{m}$, where $e$ is the charge of the particle of interest, $m$ its mass, and $c$ the speed of light. This Lagrangian is evidently in the form of a non-canonical phase-space Lagrangian, Eq.~\eqref{eq:noncanonical_ps_lagrangian}, with $z = (x, u)$ and $\vartheta = ( A + u b, 0)$.

The Euler-Lagrange equations corresponding to the guiding center Lagrangian describe the cross-field drifts and along-field motion while preserving the Hamiltonian character of the original full-orbit description. Letting $A^\dagger = A + u b$ and using index notation with Einstein summation, the guiding center Euler-Lagrange equations are
\begin{subequations}
\label{eq:gc_equations}
\begin{align}
  \left(A_{i,j}^\dagger - A_{j,i}^\dagger\right) \dot{x}^i - b_j \dot{u} - \mu \norm{B}_{,j} - \phi_{,j}& = 0, \quad j=1,2,3 
    \label{eq:gc_top} \\
  b_i \dot{x}^i - u & = 0, \label{eq:gc_bottom}
\end{align}
\end{subequations}
where indices appearing after a comma denote differentiation with respect to the corresponding coordinate. These equations contain: parallel motion along the magnetic field, the $E \times B$ drift, the $\nabla B$ drift, and the curvature drift. The polarization drift (relevant when time-dependent fields are considered) may be incorporated by including $E \times B$ velocity contributions in the definition of guiding center Lagrangian \cite{Cary_2009}.

\begin{figure}
  \includegraphics[width=0.4\textwidth]{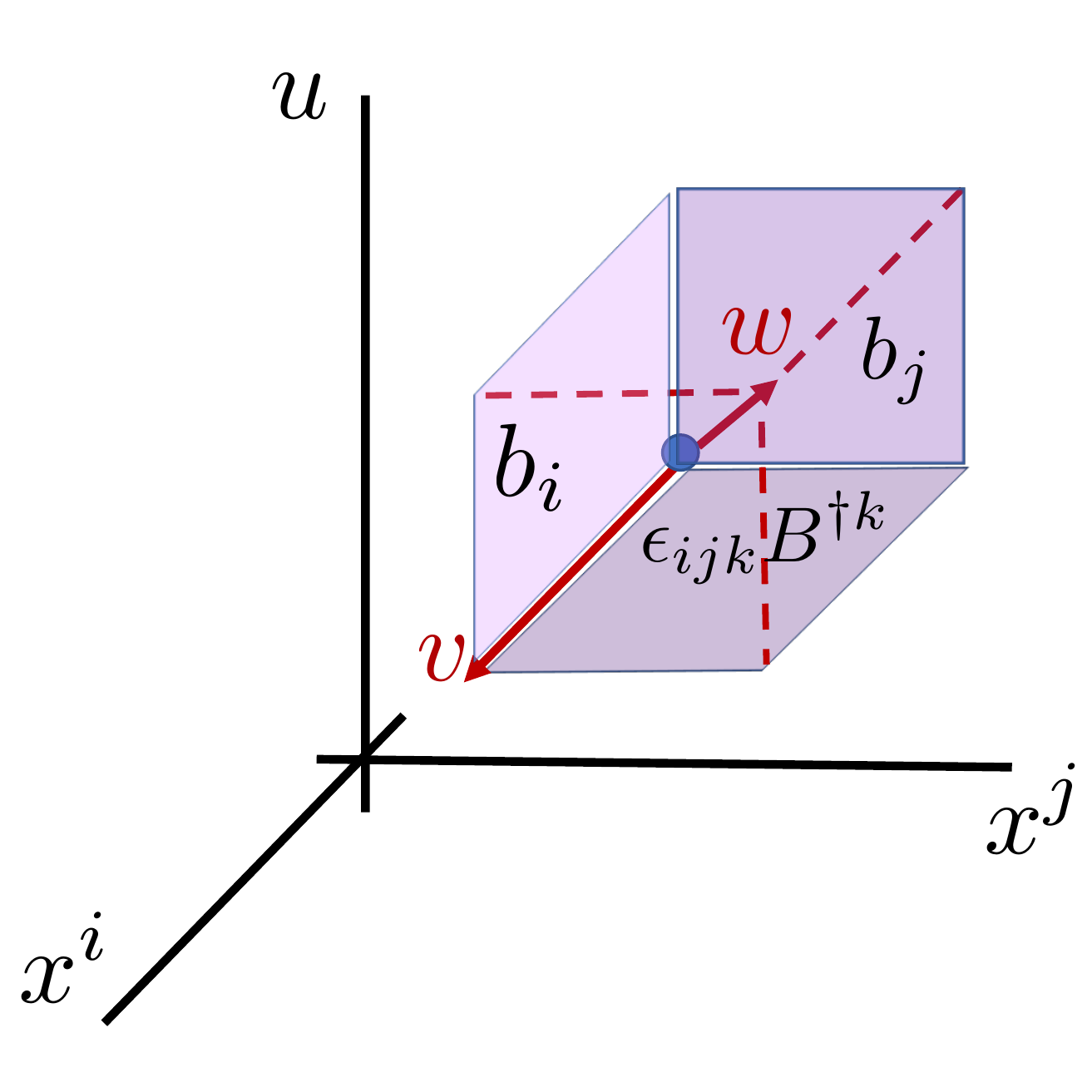}  
  \caption{Given two vectors $(v, w)$ in the $(x, u)$ guiding center phase space, the summed areas they span remains constant when weighted by the effective magnetic field $B^\dagger$ in the position planes and the magnetic field unit vector $b$ in the position-velocity planes. Here $\epsilon_{ijk}$ is the Levi-Civita symbol.}
  \label{fig:symplectic_area_guiding_center}
\end{figure}

As can be shown from the action principle in phase space, used to derive the symplectic structures in the canonical and magnetic field line settings, guiding center trajectories preserve the following symplectic structure:
\begin{equation}
  \label{eq:omega_gc}
  \Omega_{gc} = A_{i,j}^\dagger(x, u) \exteriord x^i \wedge \exteriord x^j + b_i \exteriord u \wedge \exteriord x^i.
\end{equation}
Whereas canonical Hamiltonian systems preserve areas in the $(p, q)$ phase-space plane and magnetic field lines preserve magnetic flux through the $(x^1, x^2)$ phase-space plane, guiding center trajectories preserve flux of the \emph{effective magnetic field} $B^\dagger = \nabla \times A^\dagger$ through the $(x^i, x^j)$-position coordinate planes, plus areas in the $(u, x^i)$ planes weighted by the $i^\text{th}$ component of the magnetic field unit vector. This preservation of areas weighted by the effective magnetic field and magnetic field unit vector is illustrated in Fig.~\ref{fig:symplectic_area_guiding_center}. Given two vectors $(v, w)$ in the guiding center phase space and evolving in time according to the guiding center equations, the sum of the areas they span remains constant when weighted by the indicated field components.

This four-dimensional Hamiltonian system in non-canonical coordinates $(x, u)$ comes about by performing non-canonical coordinate transformations to the charged particle phase-space Lagrangian. Early attempts at formulating the guiding center equations based on the drift dynamics alone\cite{Northrop_1963} yielded non-Hamiltonian systems. Littlejohn's seminal work on Hamiltonian guiding center theory attacked the problem by (i) beginning with the canonical phase-space Lagrangian for a particle under the influence of the Lorentz force, (ii) performing non-canonical coordinate transformations to identify the ignorable gyrophase $\theta$ and the adiabatically invariant magnetic moment $\mu$ as two of the six phase-space coordinates and (iii) truncating the guiding center Lagrangian by retaining only first-order terms in the guiding center expansion \cite{Littlejohn_1983}. Obtaining the ignorable $\theta$ and constant $\mu$ as non-canonical coordinates enabled the reduction of dimensionality from the six-dimensional full-orbit phase space to the four-dimensional (two degree-of-freedom) guiding center phase space. Meanwhile, performing all approximations upon the phase-space Lagrangian rather than the equations of motion preserved the Hamiltonian character of the system. An interesting point in the context of this work is that the degeneracy of the Lagrangian is introduced in step (i) as a means of enabling transformations on all six phase-space coordinates; the gyroaveraging/dynamical reduction does not introduce any additional degeneracy, but instead simply lowers the dimensionality of the dynamics.

Turning now to the construction of a one-step degenerate variational integrator for Eq.~\eqref{eq:gc_equations}, two assumptions will be made: First, one component of the magnetic vector potential $A_i$ will be chosen to be zero using the electromagnetic gauge freedom, as in Sec.~\ref{ssec:magnetic_field_line_flow}. Second, the \emph{same} \emph{covariant} component of the magnetic field unit vector $b_i$ will be assumed to be zero so that one component of $A^\dagger$ is eliminated altogether. This condition is satisfiable, at least with local coordinates; for instance, several of the procedures for constructing canonical coordinates for the guiding center system achieve this property as an intermediate step \cite{Zhang_2014, White_1984, White_2003}. Note however that we will \emph{not} transform to canonical coordinates when deriving this integrator. Proceeding with the assumption that, e.g., the first component of both $A$ and $b$ are zero yields a guiding center Lagrangian of the form
\begin{equation}
  \label{eq:special_gc_lagrangian}
  L(z, \dot{z}) = A_2^\dagger(z) \dot{x}^2 + A_3^\dagger(z) \dot{x}^3 - H_{gc}(z),
\end{equation}
where $z = (x, u)^T$.

In direct analogy with the discrete Lagrangian for the magnetic field line DVI, Eq.~\eqref{eq:ld_mfl_degenerate}, let us choose the following discrete Lagrangian for the guiding center dynamics:
\begin{align}
  L_d(z_k, z_{k+1}) & = L_{GC}(z_{k+1}, \frac{z_{k+1} - z_k}{h}) \nonumber \\
   & = A^\dagger(z_{k+1}) \cdot \frac{x_{k+1} - x_k}{h} - H_{gc}(z_{k+1}).
  \label{eq:ld_gc}
\end{align}
Minimizing the discrete action yields the following discrete Euler-Lagrange equations:
\begin{subequations}
 \label{eq:del_gc_original}
 \begin{align}
   \nabla A^\dagger(z_k) \cdot (x_{k} - x_{k-1}) - A^\dagger(z_{k+1}) +  A^\dagger&(z_k) -\nonumber \\
 h \nabla H_{gc}(z_k) & = 0, \\
   \nabla_u A^\dagger(z_k) \cdot (x_k - x_{k-1}) - h \nabla_u H_{gc}(z_k) & = 0.
 \end{align}
\end{subequations}
As in Section~\ref{ssec:magnetic_field_line_flow}, it appears at first glance as if a multistep scheme has been obtained; the discrete Euler-Lagrange equations include variables evaluated at times $t_{k-1}, t_k, t_{k+1}$. The Hessian informs us, however, that the chosen discrete Lagrangian is indeed properly degenerate:
\begin{align}
  \label{eq:dvi_gc_hessian}
  \frac{\partial^2 L_d(z_k, z_{k+1})}{\partial z_k \partial z_{k+1}} & = \matrixTwobyTwo{-h \nabla A^\dagger(z_{k+1})}{0}{0}{0} \nonumber \\
& = \left(\begin{array}{cccc} 0 & 0 & 0 & 0 \\ A^\dagger_{2,1} & A^\dagger_{2,2} & A^\dagger_{2,3} & 0 \\ A^\dagger_{3,1} & A^\dagger_{3,2} & A^\dagger_{3,3} & 0 \\ 0 & 0 & 0 & 0\end{array} \right),
\end{align}
where all evaluations in the final term are at $z_{k+1}$. The Hessian therefore has rank two, which is equal to the number of degrees-of-freedom of the guiding center system so that the chosen discrete Lagrangian is properly degenerate. Because the discrete Lagrangian is degenerate, the discrete Euler-Lagrange equations in Eq.~\eqref{eq:del_gc_original} cannot specify a two-step method; because it is properly degenerate, we claim (and proceed to demonstrate) that it is a one-step method. At this point, the necessity of the condition on $A$ and $b$ is apparent: if they did not share a non-zero component, then the Hessian would have non-zero terms in the top row and would not have rank two. Furthermore, because $A^\dagger$ has only two non-zero components, variables at time $t_{k+1}$ only appear in two components of Eq.~\eqref{eq:del_gc_original}, which will allow us to reformulate the equations as a one-step method. This reformulation proceeds by: (i) advancing any equations lacking a $z_{k+1}$ forward in time one index and (ii) replacing $x_k^2 - x_{k-1}^2$ and $x_k^3 - x_{k-1}^3$  in the apparently three-step equations as functions of variables at time $t_k$. The resulting one-step formulation of Eq.~\eqref{eq:del_gc_original} is
\begin{widetext}
\begin{equation}
\label{eq:dvi_gc_onestep}
\begin{split}
        A^\dagger_{2,1}(z_{k+1})(x^2_{k+1} - x^2_k) + A_{3,1}^\dagger(z_{k+1})(x^3_{k+1} - x^3_k) - h (\mu B_{,1}(x_{k+1}) + \phi_{,1}(x_{k+1})) & = 0, \\
  A_{2,2}^\dagger(z_k) \Delta^2 + A_{3,2}^\dagger(z_k) \Delta^3 - (A_2^\dagger(z_{k+1}) - A_2^\dagger(z_k)) - h (\mu B_{,2}(x_{k}) + \phi_{,2}(x_{k})) & = 0, \\
  A_{2,3}^\dagger(z_k) \Delta^2 + A_{3,3}^\dagger(z_k) \Delta^3 - (A_3^\dagger(z_{k+1}) - A_3^\dagger(z_k)) - h (\mu B_{,3}(x_{k}) + \phi_{,3}(x_{k})) & = 0, \\
b_2(x_{k+1})(x^2_{k+1} - x^2_k) + b_3(x_{k+1})(x^3_{k+1} - x^3_k) - h u_{k+1} & = 0,
\end{split}
\end{equation}
\end{widetext}
where, according to Eq.~\eqref{eq:del_gc_original}, $\Delta$ is given by
\begin{equation}
  \label{eq:delta}
  \matrixTwobyTwo{A^\dagger_{2,1}}{A^\dagger_{3,1}}{b_2}{b_3} \vectorTwo{\Delta^2}{\Delta^3} = h \vectorTwo{\mu B_{,1} + \phi_{,1}}{u_k},
\end{equation}
where all fields are evaluated at $(z_k)$. 

Being variational in nature, the guiding center DVI preserves a symplectic two-form. Restricting the discrete action to act only on trajectories generated by the one-step DVI, the derivative of the restricted discrete action is
\begin{equation}
  \exteriord \bar{S}_d(z_0) = -A^\dagger_i(z_1) \exteriord x^i_0 - A^\dagger_i(z_{N+1}) \exteriord x^i_N, 
\end{equation}
where, letting $F$ denote the one-step DVI advance (as in Eq.\eqref{eq:omega_d_mfl_pep}), $z_{k} = F^{k}(z_0)$. This equation is analogous to Eq.~\eqref{eq:dsd_mfl} for the magnetic field line problem and Eq.~\eqref{eq:dsd_canonical} for canonical systems. Taking a second exterior derivative identifies the symplectic structure preserved by the DVI to be
\begin{equation}
  \Omega_d = A_{i,j}^\dagger(F(z)) \partiald{F^j}{z^k} \exteriord z^k \wedge \exteriord x^i,
\end{equation}
which is analogous to Eq.~\eqref{eq:omega_d_mfl} in the magnetic field line problem. Because of the presence of the one-step map $F$ in this symplectic two-form, it does not immediately resemble the symplectic two-form preserved by the continuous guiding center dynamics, Eq.~\eqref{eq:omega_gc}, which may also be written:
\begin{equation}
  \Omega_{gc} = A_{i,j}^\dagger(z) \exteriord z^j \wedge \exteriord x^i,
\end{equation}
with $i=1,2,3$ and $j=(1, 2, 3, 4)$ to account for $z = (x, u)$. In this representation, it is more apparent that the guiding center DVI preserves a symplectic two-form that approaches the guiding center symplectic two-form in the zero stepsize limit. As $h \rightarrow 0$, $F$ approaches the identity map, so the symplectic two-form becomes:
\begin{equation}
 \lim_{h\to 0} \Omega_d = A^\dagger_{i,j}(x_0) \exteriord z_0^j \wedge \exteriord x_0^i, 
\end{equation}
i.e. $\lim_{h \to 0} \Omega_d = \Omega_{gc}$.

As with the magnetic field line DVI, a natural improvement to the guiding center DVI would be to obtain second-order accuracy in time, for instance by achieving a time-centered algorithm. Care must be taken that such time centering neither ruins the degeneracy (i.e., makes the method multistep) nor loses the conservation properties. Previous guiding center variational integrators used time-centered discrete Lagrangians with such accuracy in mind \cite{Qin_2008, Qin_2009, Li_2011, Kraus_2013}. However, all of these methods were not properly degenerate and therefore multistep and unstable \cite{Ellison_2015_PPCF, Ellison_thesis}. Even with the assumption on the magnetic field used in this paper, previously attempted \cite{Qin_2009, Li_2011} time centered discretizations yield multistep methods. As discussed in Sections~\ref{ssec:magnetic_field_line_flow} and \ref{ssec:canonical_hamiltonian_systems}, second-order accuracy can also be achieved by composing the above DVI with its adjoint, formed by interchanging $z_k \leftrightarrow z_{k+1}$ and mapping $h \mapsto -h$ in the discrete Lagrangian. In the canonical setting, this approach succeeds at retaining the symplectic property in the composed map. Much like the magnetic field line problem, however, the adjoint guiding center DVI does not preserve the same symplectic structure as the presented DVI. Because it is then unclear what symplectic structure might be preserved by the composition of these two methods, it may be preferable to instead use an accuracy-enhancing processing technique \cite{Blanes_2004, He_2016} rather than a time symmetrization approach; for now, this remains as future work.

Of course, one way to achieve higher-order accuracy would be to use canonical guiding center coordinates \cite{White_2003, Zhang_2014}; many high-order canonical symplectic integration schemes are known \cite{Forest_1990, Yoshida_1990, Hairer_2006_symplectic}. In fact, it is trivial to identify canonical coordinates for the guiding center Lagrangian after assuming $A_1 = b_1 = 0$. Choosing 
\begin{subequations}
\begin{align}
  p_2 & = A_2^\dagger, \\
  p_3 & = A_3^\dagger,
\end{align}
\end{subequations} 
Eq.~\eqref{eq:special_gc_lagrangian} becomes:
\begin{align}
  \label{eq:5}
  & L(x^2, x^3, p_2, p_3, \dot{x}^2, \dot{x}^3, \dot{p}_2, \dot{p}_3) = \nonumber \\
& \quad p_2 \dot{x}^2 + p_3 \dot{x}^3 - H_{gc}(x^2, x^3, p_2, p_3),
\end{align}
i.e., a canonical phase-space Lagrangian. The disadvantage to this approach is that the Hamiltonian is typically known as a function of $(x^1, x^2, x^3, u)$ without an explicit representation in the canonical coordinates. Although the transformation \emph{to} canonical coordinates is explicit, the inverse transformation to the non-canonical coordinates (in which the field functions are typically defined) requires an iterative scheme. These iterations incur computational expense; the DVI algorithm in non-canonical coordinates avoids these iterations and therefore has the prospect of being advantageous relative to canonical symplectic integration. 

A similar comparison can be made with the ``projected variational integrators'' developed in Ref.~\onlinecite{Kraus_2017_PVI}. Projected variational integrators address the problem of variational integration of phase-space Lagrangians by formulating the problem as a high dimensional canonical Hamiltonian system subject to constraints. For guiding center dynamics, the 4-D non-canonical system can be represented as an 8-D canonical system subject to (four) algebraic constraints. The projections ensure the numerical trajectory satisfies the same constraints as those governing the continuous system, and are effective for achieving good long-term behavior \cite{Kraus_2017_PVI}. Although a variational formulation for the post-projected dynamics has not been found, it has been shown to be symplectic \cite{Kraus_2017_PVI}. Advantages of projected variational integrators include their applicability to any (non-canonical) phase-space Lagrangian and their higher-order accuracy, including second- and fourth-order accuracy. The disadvantage relative to degenerate variational integration is the introduction of additional variables in the nonlinear solve, making them less efficient in this sense.

\section{Numerical Demonstrations}
\label{sec:numerical_demonstrations}

In this section, we demonstrate the long-term fidelity achieved by DVIs through their variational, structure preserving formulation. The instabilities inherent to previous variational integrators for these systems are shown to be eliminated. The benefits of the conservation properties are illustrated by showing that the DVIs capture the correct qualitative behavior of the Hamiltonian dynamics; a feature that is lost by non-conservative schemes even if they have high-order local accuracy.

The numerical examples in this section utilize tokamak magnetic geometry represented in toroidal coordinates $(r, \theta, \phi)$, where $r$ is the minor-radial position, $\theta$ the geometric poloidal angle, and $\phi$ the geometric toroidal angle. Two magnetic fields are considered: the axisymmetric magnetic field used in Ref.~\onlinecite{Qin_2009}, and the same field with an added resonant magnetic perturbation. The axisymmetric magnetic field is given by \cite{Qin_2009}
\begin{equation}
  \label{eq:axisymmetric_b}
  B(r, \theta, \phi) = \frac{B_0}{q_0(R_0 + r \cos{\theta})} e_\theta + \frac{B_0 R_0}{(R_0 + r \cos{\theta})^2} e_\phi,
\end{equation}
where $B_0$ is the on-axis magnetic field magnitude, $q_0$ the on-axis safety factor, $R_0$ the major radius and $e_\theta, e_\phi$ basis elements for contravariant vectors in toroidal coordinates (i.e., $e_\theta, e_\phi$ are not unit vectors). This axisymmetric magnetic field may be derived from the magnetic vector potential
\begin{align}
  \label{eq:axisymmetric_vector_potential}
  & A(r, \theta, \phi) = \nonumber \\
& B_0 R_0 \left( \frac{r}{\cos \theta} - \frac{R_0 \log\left(1 + \frac{r \cos \theta}{R_0}\right)}{\cos^2 \theta} \right) \nabla \theta - \frac{B_0 r^2}{2 q_0} \nabla \phi.
\end{align}
Note that $A_r = 0$, as posited in the development of the magnetic field line and guiding center variational integrators. It is also the case that $b_r=0$ for these axisymmetric fields, so the guiding center DVI can be constructed in the $(r, \theta, \phi)$ coordinates.

The second magnetic configuration applies resonant perturbations to the first vector potential:
\begin{equation}
  \label{eq:rmp_vector_potential}
  \tilde{A}(r, \theta, \phi) = A(r, \theta, \phi) - \frac{B_0 r^2}{2 q_0} \sum_i \delta_i \sin(m_i \theta - n_i \phi) \nabla \phi,
\end{equation}
where $A(r, \theta, \phi)$ is given by Eq.~\eqref{eq:axisymmetric_vector_potential}, and $\delta_i$ is the size of the $i$'th perturbation with mode numbers $m_i$ and $n_i$. This perturbation in the toroidal component of $A$ incurs perturbations in the radial and poloidal components of the otherwise axisymmetric magnetic field in Eq.~\eqref{eq:axisymmetric_b}.

\subsection{Magnetic Field Line Flow}

First, the importance of proper degeneracy is emphasized by tracing field lines in the axisymmetric configuration using the degenerate variational integrator in Eq.~\eqref{eq:mfl_dvi} and a non-degenerate (i.e., multistep) variational integrator defined by the discrete Lagrangian in Eq.~\eqref{eq:mfl_backward_ld} \emph{without} one component of the magnetic vector potential being zero. For this latter variational integrator, we use the following magnetic vector potential:
\begin{subequations}
\begin{align}
  \bar{A} & = \bar{A}_r \nabla r + \frac{A_\theta}{2} \nabla \theta + A_\phi \nabla \phi \
\\
\bar{A}_r & = -\frac{B_0 R_0 r}{\sqrt{R_0^2 - r^2}} \arctan\left(\frac{(R_0 - r) \tan(\theta/2)}
                             {\sqrt{R_0^2 - r^2}}\right),
\end{align}
\end{subequations}
where $A_\theta, A_\phi$ are defined according to Eq.~\eqref{eq:axisymmetric_vector_potential}. This is simply a gauge transformation of the original vector potential, intended to violate the condition that one component of the potential be zero. In these studies, we use $B_0 = 1$T, $R_0 = 100$cm, and $q_0 = \sqrt{2}$.

\begin{figure}
  \centering
    \includegraphics[width=0.5\textwidth]{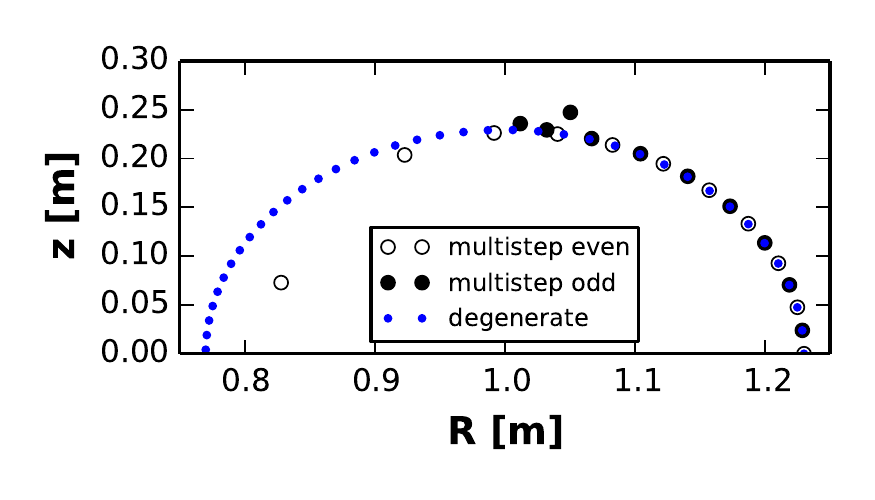}
    \caption{The multistep VI exhibits parasitic mode instabilities when used to trace magnetic field lines, wheres the DVI generates a smooth trajectory. Here we show the $(R, z)$ projection of successive steps in the trajectory, where $R = R_0 + r$.}
    \label{fig:multistep_bfield}
\end{figure}

In Fig.~\ref{fig:multistep_bfield}, a field line is traced using the DVI and the non-degenerate, multistep variational integrator. The trajectory generated by the multistep variational integrator is highly unstable; parasitic mode oscillations cause the even- and odd-indexed steps in the trajectory to diverge after just a few steps. Meanwhile, the DVI exhibits no such instability by virtue of being a one-step method.

\begin{figure}
  \centering
    \includegraphics[width=0.5\textwidth]{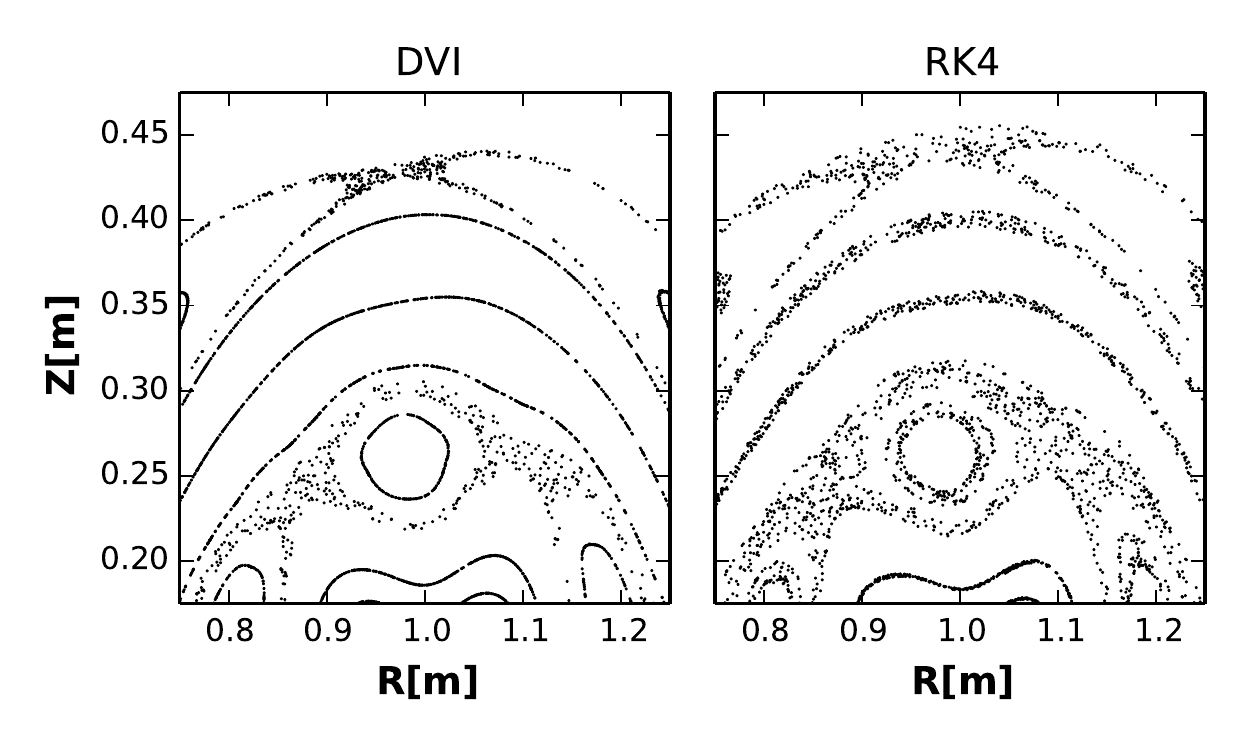}
    \caption{The magnetic field line DVI correctly distinguishes between integrable and stochastic trajectories in the resonantly perturbed tokamak, whereas the non-conservative RK4 algorithm blurs the distinction. Figure reproduced with permission from Ref.\cite{Ellison_thesis}.}
    \label{fig:mfl_poinc_portrait}
\end{figure}

The next demonstration verifies that the DVI captures the Hamiltonian nature of the magnetic field line equations. In particular, we simulate a resonantly perturbed tokamak with two Fourier components: an $(m_1=3, n_1=2)$ harmonic and an $(m_2=7, n_2=5)$ harmonic, both with amplitude $\delta_i = 3.5\times10^{-4}$, for $i=1,2$. Several field lines are initialized beginning at different radii. A Poincar\'e section is then formed by intersecting the trajectory with a plane of constant toroidal angle $\phi$. Both integrations use a step size of $h=0.5$ and advance for $3\times 10^6$ steps. The results for the DVI may be found in the left panel of Fig.~\ref{fig:mfl_poinc_portrait} and contrasted with the results of a fourth-order Runge-Kutta simulation on the right.

The salient difference between the two algorithms is that the DVI readily distinguishes between \emph{integrable} and \emph{stochastic} trajectories. Integrable trajectories correspond to those that reside on a magnetic flux surface, generating one-dimensional curves in the Poincar\'e surface of section. Meanwhile, stochastic trajectories arise near separatrices and fill two-dimensional areas in the Poincar\'e section. Of course, in the axisymmetric, $\delta_i \rightarrow 0$ limit, all magnetic field lines reside on flux surfaces. As small, symmetry-breaking perturbations are introduced, narrow stochastic regions emerge near trajectories that resonate with the perturbation. It is a Hamiltonian property, explained by the KAM theorem, that these regions should be bounded by magnetic flux surfaces for sufficiently small perturbations. This fundamentally Hamiltonian behavior is readily identifiable in the trajectory calculated using the DVI. On the other hand, although the fourth-order Runge-Kutta (RK4) scheme possesses much less local error, all of the trajectories appear area-filling in the Poincar\'e section over long integrations. Note that the numerical step size was chosen to be small enough to avoid introducing numerically induced stochasticity \cite{Friedman_1991}; the DVI portrait appears similar even at smaller step sizes. In brief, the DVI algorithm preserves the Hamiltonian character of the dynamics and is therefore well suited for the generation of Poincar\'e sections.

To best illustrate the qualitative distinctions between the algorithms, this comparison was performed at equal numerical step size. Note, however, that the implicit DVI is several times more computationally expensive, per step, than the explicit RK4 scheme. An equal computational expense comparison, allowing RK4 to execute more steps of smaller size, would eventually reveal the same qualitative behavior but requires much longer simulations.

\subsection{Guiding Center Trajectories}

The non-canonical guiding center system bares many similarities to the non-canonical magnetic field line system. Here we show that the benefits demonstrated in the magnetic field line context carry over to the guiding center system. 

\begin{figure}
  \includegraphics[width=0.5\textwidth]{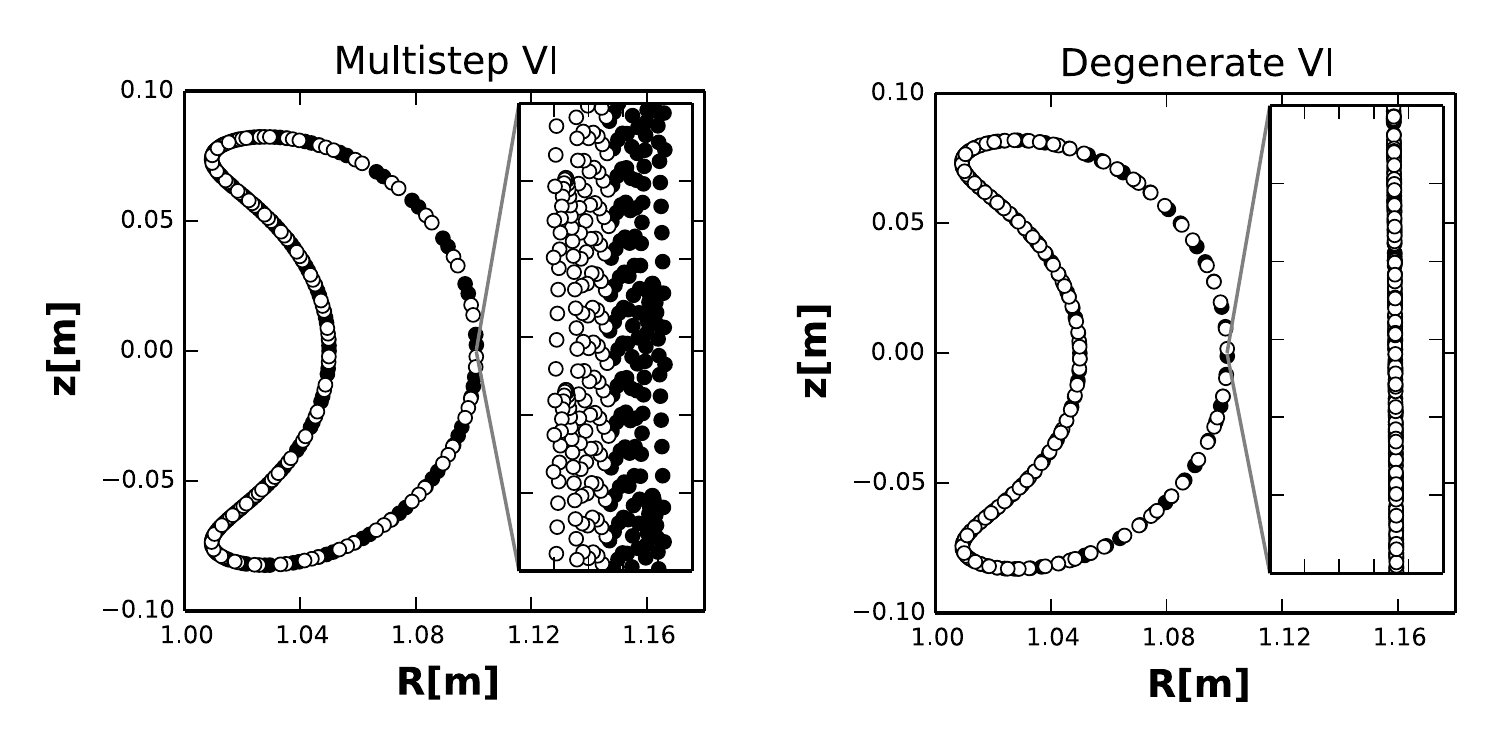}
  \caption{The guiding center DVI successfully eliminates the parasitic mode instabilities present in existing variational guiding center algorithms. Even-indexed points are marked in white, and odd-indexed points in black. These trapped particle trajectories were evolved in the axisymmetric magnetic field given in Eq.~\eqref{eq:axisymmetric_b}. The insets have a width of $10^{-4}$ m in the $R$-direction and $3 \times 10^{-4}$ m in the $z$-direction. Although the parasitic mode oscillations appear small, they cause the Newton-Rhapson iterations to fail to converge, crashing the simulation. Figure reproduced with permission from Ref.\cite{Ellison_thesis}.}
  \label{fig:gc_parasitic}
\end{figure}

Preceding this work, guiding center variational integrators have been multistep methods \cite{Qin_2008, Qin_2009, Li_2011, Kraus_2013}. In certain configurations, the parasitic modes inherent to these methods can become unstable, leading to even-odd oscillations akin to those observed in Figs.~\ref{fig:parasitic_pendulum} and \ref{fig:multistep_bfield}. As an example of this behavior in the guiding center context, we first produce a guiding center trapped-particle ``banana orbit'' trajectory using the variational integrator developed in Ref.~\onlinecite{Li_2011}, which uses the discrete Lagrangian
\small
\begin{align}
  &L_d(x_k, x_{k+1}, u_{k+1/2}) = \nonumber \\ &A^\dagger(\frac{x_k + x_{k+1}}{2}, u_{k+1/2}) \cdot \frac{x_{k+1} - x_k}{h} -
 h H_{gc}(\frac{x_k + x_{k+1}}{2}, u_{k+1/2}).
  \label{eq:ld_gc_staggered_midpoint}
\end{align}
\normalsize
Here, the parallel velocity coordinate $u$ has been ``staggered'' in time with respect to the position coordinates. The staggering of this coordinate is made possible by the absence of its time derivative, $\dot{u}$, in the Lagrangian. The motivation for such a staggering was to enhance the stability of the algorithm\cite{Li_2011}, which we now understand to be related to the parasitic mode oscillations. Indeed, the algorithm was believed to be stable when first presented; the parasitic modes are difficult to detect for many configurations, depending on the electric and magnetic fields, the particle's initial condition, and the numerical step size. As evidenced by the left panel of Fig.~\ref{fig:gc_parasitic}, however, parasitic modes remain present in this variational integrator and are unstable under certain conditions. The figure depicts a trapped particle ``banana-orbit'' trajectory in the axisymmetric magnetic field in Eq.~\eqref{eq:axisymmetric_b} with $B_0 = 1$ T, $R_0=100$ cm, $q_0 = \sqrt{2}$ and with initial condition $(r, \theta, \phi, u, \mu) = (5, 0, 0, -0.129, 2.1\times 10^{-4})$ in the normalized units. In this and all following figures, the electric field $E = -\nabla \phi - \partiald{A}{t}$ is taken to be zero. The numerical step size corresponds to roughly 600 steps per bounce period. At early times, the even- and odd-indexed steps form a smooth trajectory in the $R, z$-plane. These are the central points in the inset zoom. As time progresses, however, the even- and odd-indexed trajectories diverge, leading to the leftmost white and rightmost black markers, respectively. Soon after the depicted time, the nonlinear solve fails to converge due to the large amplitude of the parasitic mode oscillation.

The multistep variational integrator tested above requires supplying excess initial conditions and admits parasitic mode oscillations. Interestingly, the staggering of the parallel velocity coordinate introduces some of the desirable degeneracy in the discrete Lagrangian, albeit not enough to obtain a one-step method and eliminate all of the parasitic modes. The integrator specifies an update from $(x_{k-1}, u_{k-1/2}, x_{k})$ to $(u_{k+1/2}, x_{k+1})$, so two position initial conditions are required but only one parallel velocity initial condition is required. From the perspective of a degeneracy calculation, this manifests as the Hessian matrix for the discrete Lagrangian in Eq.~\eqref{eq:ld_gc_staggered_midpoint} having rank three. Although it is degenerate (full rank would be four), it is not \emph{properly degenerate}, which would be rank two for the guiding center system.

In Section~\ref{ssec:guiding_center_trajectories}, we were able to construct a discrete Lagrangian with a rank two Hessian by using a non-centered time discretization and assuming that the magnetic vector potential and magnetic field unit vector had a common component that was everywhere zero. The axisymmetric magnetic field employed in this test case satisfies these assumptions. The properly degenerate discrete Lagrangian led to the DVI in Eq.~\eqref{eq:dvi_gc_onestep}. In the right panel of Fig.~\ref{fig:gc_parasitic}, we evolve the trapped particle trajectory using said DVI under the same conditions as those used for the multistep variational integrator. The smooth trajectory is evidence that the parasitic modes have been completely eliminated. Because the DVI is a one-step method, no parasitic modes can be present. 

\begin{figure}
  \centering
    \includegraphics[width=0.35\textwidth]{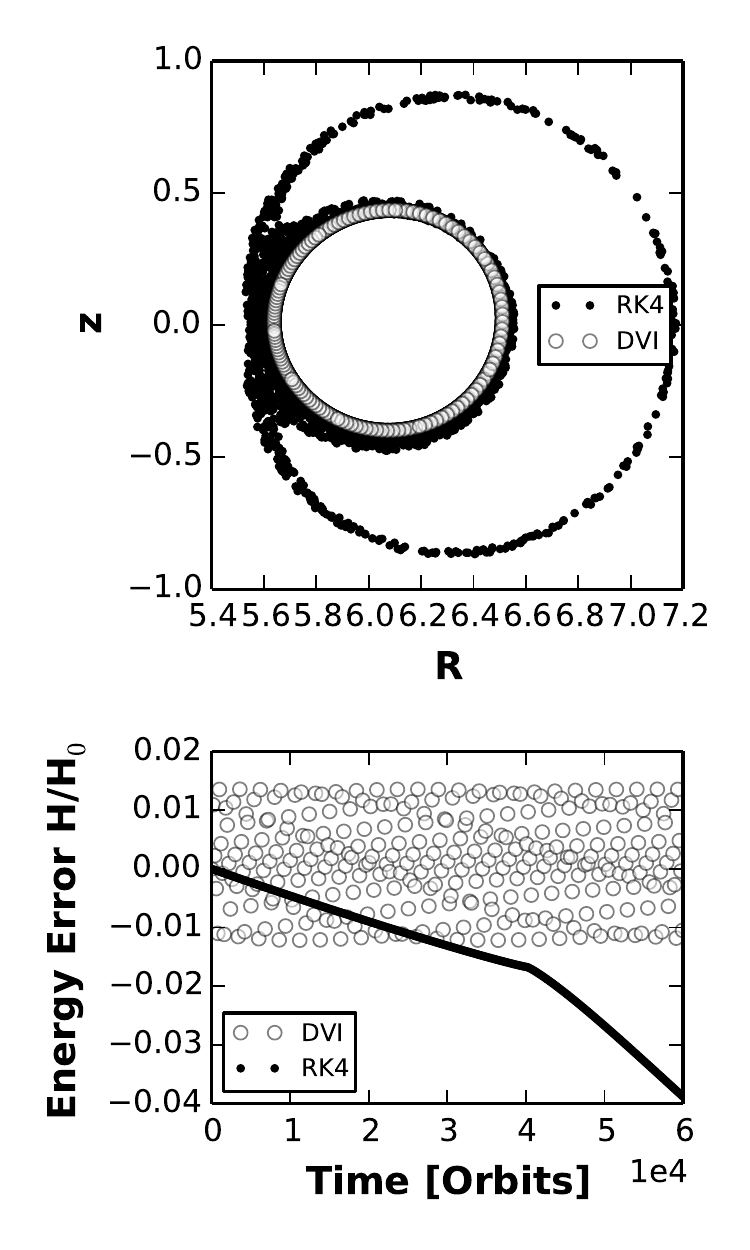}
    \caption{The DVI exhibits excellent long-term fidelity in the integration of a 3.5 MeV passing alpha-particle orbit. The RK4 advance accumulates global errors in an undesirable way, eventually leading to the transition to a trapped particle banana orbit.}
    \label{fig:gc_passing_dvi_rk4}
\end{figure}

Now that the stability of the guiding center DVI has been established, we turn to demonstrating its long-term fidelity imbued by its preservation of a non-canonical symplectic structure. In Fig.~\ref{fig:gc_passing_dvi_rk4}, a passing particle trajectory is advanced using the DVI algorithm and RK4. Parameters for this study were chosen to correspond to a 3.5 MeV alpha particle in an ITER-like configuration. In the normalized units, the field parameters are given by: $B_0 = 255.6$, $R_0=6.20$, $q_0 = \sqrt{2}$ and the particle's initial condition was $(r, \theta, \phi, u, \mu) = (0.31, 0, 0, -5.2, 0.277)$ \footnote{Recall that $B$ is normalized by $e/mc$.}. Equal numerical step sizes were used for the two algorithms with 25 steps per orbit period (a complete revolution in the $R-z$ plane). 

Although the RK4 algorithm introduces much less \emph{local} truncation error, the truncation errors accumulate in qualitatively different manners for the two schemes. The energy error for RK4 is unbounded and decreasing; meanwhile, the energy error for the DVI algorithm remains bounded indefinitely (with relatively large oscillations due to its first-order accuracy). Such behavior is indicative of the DVI preserving \emph{some} energy function that differs from the true energy in a stepsize-dependent way (c.f. Eq.~\eqref{eq:modified_hamiltonian}). The differences between the schemes are also apparent in the evolution of the trajectory in the $R-z$ plane: the DVI maintains a closed trajectory, whereas RK4 exhibits an increasingly distorted trajectory, eventually transitioning from a passing particle orbit to a trapped particle orbit. Such unphysical transitions in orbit characteristics are clearly undesirable when modeling energetic particle processes.

The preceding comparison was performed at equal numerical step size to highlight the qualitative distinctions in the trajectories generated by the respective algorithms. Which algorithm proves preferable for a specific application depends on many factors including: the required accuracy, the numerical quantities of interest, the timescales of interest, and the relative computational expense of the two algorithms. Over sufficiently long times, the conservative algorithm will eventually out-perform the non-conservative algorithm, but the timescale for this may be longer than the timescale of interest for particular studies especially under equal computational expense comparisons; the implicit DVI advance is several times more expensive than the explicit RK4 advance on a per-step basis \cite{Ellison_thesis}, and the error of the RK4 scheme, being fourth-order accurate, decreases rapidly as the step size is reduced. Still, energetic particle processes often require long time integrations, and there is great interest in conservative algorithms for modeling these processes \cite{Qin_2008, Qin_2009, Li_2011, Kraus_2013, He_2015, He_2016, Wang_2017}. The conservative character of the DVI does not preclude the incorporation of dissipative dynamics, including collisional drag. It has been shown how to use a modified statement of the least action principle, known as a Lagrange-d'Alembert principle, to incorporate dissipative effects into the time advance \cite{Ellison_thesis}, in which case it remains important that all dissipation is due to physical effects rather than an unknown combination of physical and numerical dissipation. 

\section{Discussion}
\label{sec:discussion}

In this paper, we presented a new technique for the symplectic integration of non-canonical Hamiltonian systems. The degenerate variational integrator method focuses on discretizing the phase-space action principles that play a prominent role in plasma physics. We have shown that in order to capture the geometry of the dynamics --- including the order of the dynamical system and the area-preserving symplectic flow --- it is important that the variational discretization retain the degeneracy of the Lagrangian. Toward this end, we provided a simple means of checking for degeneracy based on a Hessian matrix of second-order derivatives. 

For canonical Hamiltonian systems, the DVI technique can be used to derive the familiar leapfrog advance. For non-canonical applications of interest to  plasma physics, the DVI technique enables the first one-step non-canonical symplectic integrators for magnetic field line flow and a class of guiding center trajectories. In both cases, an electromagnetic gauge is chosen to facilitate the degenerate discretization. For guiding center trajectories, an additional assumption about the orientation of the magnetic field is also assumed (an assumption relaxed in a related work \cite{Burby_2017}). The numerical demonstrations presented here verify the new, non-canonical symplectic integrators capture the qualitative Hamiltonian behavior of the dynamics. Practical, equal-expense comparisons suggest it is favorable to use the guiding center DVI for energetic particle processes. 

Prospects for future work persist in both developing advanced algorithms for modeling plasma systems and in the numerical analysis of variational integrators for degenerate Lagrangian systems. In terms of plasma physics algorithms, natural progressions of this work include achieving higher-order accuracy and adaptive time stepping. While pursuing higher-order accuracy, the care must be taken to retain the preservation of the non-canonical symplectic structure. Eventually, this particle-advance scheme and its decedents may be used in structure-preserving drift- and gyro-kinetic simulations, analogous to recent work in multisymplectic PIC simulations \cite{Squire_2012_PIC, Qin_2016, Kraus_2017}. In terms of the numerical analysis of variational integrators, much remains to be explored for degenerate variational integrators. For instance, the non-canonical symplectic structures that converge to the continuous structure in the zero-step size limit are rather unique. Much of the analysis of canonical symplectic integrators assumes the algorithm preserves the same two-form; it will be valuable and interesting to determine rigorous implications of this more general class of symplectic algorithms. Additionally, it would be important to determine how to construct a properly degenerate variational integrator for the fully general non-canonical phase-space Lagrangian in Eq.~\eqref{eq:noncanonical_ps_lagrangian}, thereby providing a solution for the longstanding issue of symplectic integration in non-canonical coordinates.

\section{Acknowledgments}

This work was performed under the auspices of the U.S. Department of Energy by Lawrence Livermore National Laboratory under contract DE-AC52-07NA27344; under the auspices of the NNSA of the U. S. DOE by Los Alamos National Laboratory, operated by LANS LLC under DOE contract No. DEAC52-06NA25396; and by the Princeton Plasma Physics Laboratory under contract number DE-AC02-09CH11466. This project has received funding from the European Union’s Horizon 2020 research and innovation programme under the Marie Sk{\l}odowska-Curie grant agreement No 708124. The authors are grateful to Jonathan Squire, Yao Zhou and Daniel Ruiz for valuable conversations. LLNL-JRNL-744319-DRAFT

\appendix 

\section{Differential Geometry Primer}
\label{sec:differential_geometry_primer}

\newcommand{\myvec}[1]{\mathbf{#1}}

Differential geometry and, in particular, the theory of differential forms, generalizes the operations of tensor calculus (including $\divergence, \grad, \curl$) to higher dimensional spaces and to manifolds. Differential forms inhabit a central role in classical mechanics, illuminating relationships that would be tedious and opaque if they were to be expressed in more traditional tensor calculus notation. For a detailed discussion, the interested reader is directed to one of the many references on the subject, e.g. Refs.~\onlinecite{Holm_2009, Frankel_2012, Marsden_1999}. This appendix is intended to rapidly familiarize the reader with the basics of differential forms as employed in the text. 

Consider an $n$-dimensional space $\reals^n$ with arbitrary (e.g., curvilinear) local coordinates $z^1, z^2, ..., z^n$. At each point $\myvec{p}$ in this space, there exists a collection of vectors tangent to the space at the point $\myvec{p}$. In physics parlance, these are referred to as ``contravariant vectors''; in the language of differential geometry, they are deemed ``tangent vectors''. Basis elements for this vector space may be chosen to be variations of the position $\myvec{p}$ with respect to the coordinate functions:
\begin{equation}
  \label{eq:tangent_basis}
  (\myvec{e}_1, ..., \myvec{e}_n) = (\partiald{\myvec{p}}{z^1}, ..., \partiald{\myvec{p}}{z^n}).
\end{equation}
For shorthand, the $i$-th basis element will hereafter be denoted $\partiald{}{z^i}$. So, any vector $\myvec{v}$ tangent to the point $\myvec{p}$ may be written with respect to this basis:
\begin{equation}
  \label{eq:tangent_vector}
  \myvec{v} = v^i \partiald{}{z^i},
\end{equation}
where Einstein summation convention over repeated sub- and super-scripted indices is implied.

Next, consider a smooth function $f$ defined on this space with $f(\myvec{p}) \in \reals$ for each $\myvec{p}$ in the space. The \emph{differential} of $f$, $\exteriord f$, operates on tangent vectors to yield real numbers:
\begin{equation}
  \label{eq:differential_f}
  \exteriord f(\myvec{p}) \cdot \myvec{v} = \partiald{f}{z^i}(\myvec{p})  v^i.
\end{equation}
The differential of $f$ acting on $\myvec{v}$ is simply the directional derivative of $f$. Although the above operation was notated with ``$\cdot$'', this differs from the vector dot product in that \emph{no metric tensor was required}. Because $\exteriord f$ can be combined with vectors in such a manner, the dependence on the position $\myvec{p}$ will no longer be explicitly notated, preferring to reserve the argument for vectors on which $\exteriord f$ might act, for instance $\exteriord f (\myvec{v})$. 

Notice that $\exteriord f(\myvec{v})$ is a linear function on vectors $\myvec{v}$ tangent to $\reals^n$ at $\myvec{p}$, so $\exteriord f$ must be a member of the \emph{dual} of the space to which $\myvec{v}$ is a member. In differential geometry, the dual of a vector is called a \emph{1-form} (as opposed to a covariant vector). The basis elements for 1-forms are simply the differentials of the coordinate functions
\begin{equation}
  (\myvec{e}^1, ..., \myvec{e}^n) = (\exteriord z^1, ..., \exteriord z^n),
\end{equation}
which may be confirmed using Eq.~\eqref{eq:differential_f} with $f = z^i$ and $v$ chosen to have a single non-zero component $v^j = 1$. The result is:
\begin{equation}
  \exteriord z^i \cdot \partiald{}{z^j} = \delta^i_j,
\end{equation}
where the Kronecker delta $\delta^i_j$ is one if $i = j$ and is otherwise zero. 

A general 1-form at some position $\myvec{p}$ then takes the form
\begin{equation}
  \alpha = a_i \exteriord z^i.
\end{equation}
In this notation, one would write the one-form corresponding to $\exteriord f$ as
\begin{equation}
  \label{eq:exteriord_f}
  \exteriord f = \partiald{f}{z^i} \exteriord z^i.
\end{equation}
This expression is in agreement with the elementary calculus interpretation of $\exteriord$ indicating an infinitesimal element; however, it is now laden with additional meaning through its relation to tangent vectors and some additional properties that we shall soon define. Note that although $\exteriord f$ is a one-form, not all one-forms are differentials of some function.

Equipped with tangent vectors and one-forms, the next level of complexity involves higher-degree tensors. Higher-rank tensors may be constructed from lower-rank tensors using the tensor product. For instance, if $\alpha$ and $\beta$ are 1-forms, a \emph{twice-covariant tensor field} is given by:
\begin{equation}
  \alpha \otimes \beta.
\end{equation}
At each point $\myvec{p}$, $\alpha \otimes \beta$ gives a bilinear function operating on pairs of tangent vectors to output a real number. For instance, if $\myvec{v}, \myvec{w}$ are tangent vectors at $\myvec{p}$, then
\begin{equation}
  \alpha \otimes \beta (\myvec{v}, \myvec{w}) = \alpha_i v^i \beta_j w^j,
\end{equation}
where it is implied that $\alpha$ and $\beta$ have been evaluated at $\myvec{p}$. Bilinearity of the tensor product of $\alpha$ and $\beta$ follows from the linearity of each of the 1-forms individually.

Anti-symmetry appears in many tensor calculus operations, including cross products and Jacobian determinants. It appears so often that an anti-symmetric version of the tensor product is the central product operation used in differential geometry. The \emph{wedge product} constructs twice-covariant anti-symmetric tensors according to:
\begin{equation}
  \label{eq:wedge_product}
  \alpha \wedge \beta = \alpha \otimes \beta - \beta \otimes \alpha.
\end{equation}
A field identifying an anti-symmetric twice-covariant tensor at each position $\myvec{p}$ in $\reals$ is known as a \emph{2-form}. A prominent 2-form appearing in this manuscript is the \emph{canonical symplectic structure} on $\reals^2$-space with coordinates $(q, p)$:
\begin{equation}
  \Omega = \exteriord q \wedge \exteriord p.
\end{equation}
At this point, we can verify that $\Omega$ determines the area spanned by two vectors at a particular point in the $(q, p)$-plane. Let $\myvec{v} = v^q \partiald{}{q} + v^p \partiald{}{p}$ and $\myvec{w} = w^q \partiald{}{q} + w^p \partiald{}{p}$, then
\begin{align*}
 \exteriord q \wedge \exteriord p (\myvec{v}, \myvec{w}) & = \exteriord q \otimes \exteriord p (\myvec{v}, \myvec{w}) - \exteriord p \otimes \exteriord q (\myvec{v}, \myvec{w})\\
& = \exteriord q (\myvec{v}) \otimes \exteriord p (\myvec{w}) - \exteriord p(\myvec{v}) \otimes \exteriord q(\myvec{w}) \\
& = v^q w^p - v^p w^q.
\end{align*}
The final expression is immediately recognizable as the area spanned by the vectors $\myvec{v}$ and $\myvec{w}$. 

The wedge product is closely related to the vector cross product, but is geometrically quite distinct. Whereas the cross product maps a pair of vectors into a vector, the wedge product maps a pair of 1-forms into a 2-form. Two-forms may be visualized as two-dimensional areas, often appearing as the integrands of surface integrals; the same cannot be said of vectors constructed using the cross product. Finally, whereas the cross product involves the metric tensor in curvilinear coordinates, \emph{no metric tensor is necessary} for either the wedge product or to apply a 2-form to tangent vectors. For instance, with polar coordinates $(r, \theta)$, 
\begin{equation}
\exteriord r \wedge \exteriord \theta (\partiald{}{r}, \partiald{}{\theta}) = 1,
\end{equation}
even though $\exteriord \theta$ and $\partiald{}{\theta}$ are not unit vectors. 

The wedge product further allows us to construct differential $k$-forms, which are multilinear, anti-symmetric, $k$-times covariant tensor fields on $\reals^n$. Just like 1-forms operate on individual vectors and 2-forms operate on pairs of vectors, $k$-forms operate on sets of $k$-vectors linearly in each argument. Anti-symmetry indicates that interchanging any two arguments incurs a minus sign; if $\alpha$ is a $k$-form, then
\begin{equation}
  \alpha(\myvec{v_1}, ..., \myvec{v_i}, ..., \myvec{v_j}, ... \myvec{v_k}) = - \alpha(\myvec{v_1}, ..., \myvec{v_j}, ..., \myvec{v_i}, ... \myvec{v_k})
\end{equation}
for any distinct $i, j \in (1, ..., k)$. This property is clearly manifested by 2-forms, as evidenced by Eq.~\eqref{eq:wedge_product}. An arbitrary $k$-form on $\reals^n$ is a linear combination of terms of the form $a(\myvec{p}) \exteriord z^{i_1} \wedge ... \wedge \exteriord z^{i_k}$.

The final operation necessary in this context is the \emph{exterior derivative}. The exterior derivative takes $k$-forms to $(k+1)$-forms and may be defined according to the following to properties. (i) The exterior derivative of a smooth function $f$ is simply the differential of $f$ (see Eq.~\eqref{eq:exteriord_f}). (ii) If $\alpha$ is a $k$-form given by:
\begin{equation}
  \alpha = a_{i_1 ... i_n} \exteriord z^{i_1} \wedge ... \wedge \exteriord z^{i_n},  
\end{equation}
then the exterior derivative of $\alpha$ is the $(k+1)$-form
\begin{equation}
  \label{eq:exteriord_operation}
  \exteriord \alpha = \partiald{a_{i_1 ... i_n}}{z^j} \exteriord z^j \wedge \exteriord z^{i_1} \wedge ... \wedge \exteriord z^{i_n}.
\end{equation}
Here the summation is over the $j$ index, but $i_1$ etc. are specific indices with no summation implied. As a simple example, the canonical symplectic structure may be recovered using the exterior derivative of a 1-form: $\Omega = \exteriord ( q \exteriord p)$.

Perhaps the most important --- and at first glance, mysterious --- property of the exterior derivative for this manuscript is that $\exteriord ( \exteriord \alpha ) = 0$ for \emph{any} $k-$form $\alpha$. This follows in a straightforward manner from the anti-symmetry of the wedge product and the equivalence of mixed partial derivatives. Demonstrating this claim in detail for a 1-form $\alpha$:
\begin{align*}
  \alpha(\myvec{z}) & = a_i(\myvec{z}) \exteriord z^i \\
  \exteriord \alpha & = a_{i,j}(\myvec{z}) \exteriord z^j \wedge \exteriord z^i,
\end{align*}
where $a_{i,j}$ denotes $\partiald{a_{i}}{z^j}$. Taking a second exterior derivative then:
\begin{align*}
  \exteriord^2 \alpha & = \exteriord( \exteriord \alpha) = a_{i, jk}(\myvec{z}) \exteriord z^k \wedge \exteriord z^j \wedge \exteriord z^i \\
  & = \half a_{i, jk}(\myvec{z}) \exteriord z^k \wedge \exteriord z^j \wedge \exteriord z^i - \half a_{i, jk}(\myvec{z}) \exteriord z^j \wedge \exteriord z^k \wedge \exteriord z^i \\
  & = \half \left(a_{i, jk} - a_{i, kj}\right) \exteriord z^k \wedge \exteriord z^j \wedge \exteriord z^i = 0,
\end{align*}
where the final equality follows from the equivalence of mixed partial derivatives. Applying $\exteriord^2$ to 1-forms therefore bears a strong resemblance to taking the curl of a gradient. The interested reader may extend the case to 2-forms and will be strongly reminded of a similar exercise taking the divergence of a curl. Unlike $\divergence, \grad$, and $\curl$, no metric tensor elements appear in the exterior derivative operations. Also unlike $\divergence, \grad, \curl$, the exterior derivative and wedge product operations extend straightforwardly beyond three dimensions. For additional details on the relationship between differential geometry and vector calculus, consult the references \cite{Holm_2009, Frankel_2012, Marsden_1999}.

\section{A degenerate Lagrangian -- Fermat's principle}
\label{sec:fermats_principle}

Degenerate Lagrangians appear in physics outside of phase-space Lagrangians. One insightful example is Fermat's principle.

Fermat's principle in two dimensions has rays with $ds/dt=1/n(x,y)$,
where $n(x,y)$ is the index of refraction, $s$ is arclength and
the speed of light equals unity. One minimizes 
\begin{equation}
T=\int n(x,y)ds=\int n(x,y)\sqrt{dx^{2}+dy^{2}}.\label{eq:Fermat-action}
\end{equation}
Letting time $t$ be the independent variable we have

\[
T=\int Ldt=\int n(x,y)\sqrt{\dot{x}^{2}+\dot{y}^{2}}dt.
\]
This is a special case of geodesics, where one seeks to minimize the
length $T_{g}=\int\sqrt{g_{ij}\dot{x}^{i}\dot{x}^{j}}dt$ with the
metric tensor $g$. As in the magnetic field line problem discussed
in Sec.~\ref{ssec:magnetic_field_line_flow}, the action ($T$ or $T_{g}$) is invariant
under time reparameterization. This property also holds for geodesics.

The Euler-Lagrange equations lead to
\begin{equation}
\left[\begin{array}{cc}
\dot{y} & -\dot{x}\\
-\dot{y} & \dot{x}
\end{array}\right]\left(\begin{array}{c}
\ddot{x}\\
\ddot{y}
\end{array}\right)=\left(\begin{array}{c}
\left(\dot{x}^{2}+\dot{y}^{2}\right)\left(\lambda_{x}\dot{y}-\lambda_{y}\dot{x}\right)\\
-\left(\dot{x}^{2}+\dot{y}^{2}\right)\left(\lambda_{x}\dot{y}-\lambda_{y}\dot{x}\right)
\end{array}\right),\label{eq:2X2-for-2nd-derivs}
\end{equation}
where $\lambda_{x}=n^{-1}\partial n/\partial x$ and $\lambda_{y}=n^{-1}\partial n/\partial y$.
The matrix here is proportional to the Hessian $\mathsf{H}$,
with $\mathsf{H}_{ij}=\partial^{2}L/\partial\dot{x}_{i}\partial\dot{x}_{j}$.
This matrix is singular, i.e.~the Lagrangian $L$ is degenerate,
and $\mathsf{H}$ is in fact of rank unity. But notice this: the vector
on the right in Eq.~(\ref{eq:2X2-for-2nd-derivs}) \emph{is in the
range} of the Hessian. For a full rank Hessian, the system is fourth
order, but with rank unity, the equations are of second order. (Also,
since the Hessian is singular, it is not possible to perform a Legendre
transformation to go to a Hamiltonian prescription.)

Alternatively, we can make $y$ the independent the `time-like' variable,
as in Sec.~\ref{ssec:magnetic_field_line_flow}:
\begin{equation}
S=\int n(x,y)\sqrt{1+(dx/dy)^{2}}dy.\label{eq:Geometric-x(y)}
\end{equation}
 The Euler-Lagrange equation leads to
\begin{equation}
\frac{d^{2}x}{dy^{2}}-\left[1+\left(\frac{dx}{dy}\right)^{2}\right]\left(\lambda_{x}-\frac{dx}{dy}\lambda_{y}\right)=0,\label{eq:Time-reparam-eq}
\end{equation}
a single second order equation. (As in Sec.~\ref{ssec:magnetic_field_line_flow},
if the $y-$variable approaches a maximum or minimum ($dx/dy\rightarrow\pm\infty$),
we must reparameterize with, for example, $x$ becoming the independent
variable.) Substituting $\dot{x}=\dot{y}dx/dy$ and $\ddot{x}=\dot{y}^{2}d^{2}x/dy^{2}+\ddot{y}dx/dy$
into either of the two equations in Eqs.~(\ref{eq:2X2-for-2nd-derivs}),
we find Eq.~(\ref{eq:Time-reparam-eq}). These equations provide
the geometric path; the rate along the path requires (because of the
time reparameterization) the use of the original relation $ds/dt=1/n(x,y)$ 

A final comment is that the `affine Fermat's principle', a special
case of affine geodesics, minimizes $T_{a}=\int L_{a}dt=(1/2)\int n(x,y)^{2}\left(\dot{x}^{2}+\dot{y}^{2}\right)dt$.
(The general `affine geodesic' minimizes $(1/2)\int g_{\mu\nu}\dot{x}_{\mu}\dot{x}_{\nu}dt$.)
The Euler-Lagrange equations for $T_{a}$ are
\begin{equation}
\ddot{x}=-\lambda_{x}\dot{x}^{2}+\lambda_{x}\dot{y}^{2}-2\lambda_{y}\dot{x}\dot{y},\label{eq:Affine1}
\end{equation}
\begin{equation}
\ddot{y}=-\lambda_{y}\dot{y}^{2}+\lambda_{y}\dot{x}^{2}-2\lambda_{x}\dot{x}\dot{y},\label{eq:Affine2}
\end{equation}
a fourth order system. This is expected, since $L_{a}$ is nondegenerate.
These equations from the affine system are consistent with either
equation of Eq.~(\ref{eq:2X2-for-2nd-derivs}) in the sense that
combining these equations to form $\dot{y}\ddot{x}-\dot{x}\ddot{y}$
indeeds leads to Eq.~(\ref{eq:2X2-for-2nd-derivs}). These affine
equations provide the path \emph{and} the rate along the path, without
needing to re-use $ds/dt=1/n(x,y)$. In fact, the Hamiltonian for
$T_{a}$ has $p_{x}=n^{2}\dot{x}$, $p_{y}=n^{2}\dot{y}$ and $H=(p_{x}^{2}+p_{y}^{2})/2n^{2}=n(x,y)^{2}\left(\dot{x}^{2}+\dot{y}^{2}\right)/2$,
so $dH/dt=0$ implies $dL_{a}/dt=0$, or $ds/dt=1/n(x,y)$.

\bibliographystyle{aipnum4-1}
\bibliography{SI_refs}

\begin{thebibliography}{57}%
\makeatletter
\providecommand \@ifxundefined [1]{%
 \@ifx{#1\undefined}
}%
\providecommand \@ifnum [1]{%
 \ifnum #1\expandafter \@firstoftwo
 \else \expandafter \@secondoftwo
 \fi
}%
\providecommand \@ifx [1]{%
 \ifx #1\expandafter \@firstoftwo
 \else \expandafter \@secondoftwo
 \fi
}%
\providecommand \natexlab [1]{#1}%
\providecommand \enquote  [1]{``#1''}%
\providecommand \bibnamefont  [1]{#1}%
\providecommand \bibfnamefont [1]{#1}%
\providecommand \citenamefont [1]{#1}%
\providecommand \href@noop [0]{\@secondoftwo}%
\providecommand \href [0]{\begingroup \@sanitize@url \@href}%
\providecommand \@href[1]{\@@startlink{#1}\@@href}%
\providecommand \@@href[1]{\endgroup#1\@@endlink}%
\providecommand \@sanitize@url [0]{\catcode `\\12\catcode `\$12\catcode
  `\&12\catcode `\#12\catcode `\^12\catcode `\_12\catcode `\%12\relax}%
\providecommand \@@startlink[1]{}%
\providecommand \@@endlink[0]{}%
\providecommand \url  [0]{\begingroup\@sanitize@url \@url }%
\providecommand \@url [1]{\endgroup\@href {#1}{\urlprefix }}%
\providecommand \urlprefix  [0]{URL }%
\providecommand \Eprint [0]{\href }%
\providecommand \doibase [0]{http://dx.doi.org/}%
\providecommand \selectlanguage [0]{\@gobble}%
\providecommand \bibinfo  [0]{\@secondoftwo}%
\providecommand \bibfield  [0]{\@secondoftwo}%
\providecommand \translation [1]{[#1]}%
\providecommand \BibitemOpen [0]{}%
\providecommand \bibitemStop [0]{}%
\providecommand \bibitemNoStop [0]{.\EOS\space}%
\providecommand \EOS [0]{\spacefactor3000\relax}%
\providecommand \BibitemShut  [1]{\csname bibitem#1\endcsname}%
\let\auto@bib@innerbib\@empty
\bibitem [{\citenamefont {Littlejohn}(1983)}]{Littlejohn_1983}%
  \BibitemOpen
  \bibfield  {author} {\bibinfo {author} {\bibfnamefont {R.~G.}\ \bibnamefont
  {Littlejohn}},\ }\href@noop {} {\bibfield  {journal} {\bibinfo  {journal}
  {Journal of Plasma Physics}\ }\textbf {\bibinfo {volume} {29}},\ \bibinfo
  {pages} {111} (\bibinfo {year} {1983})}\BibitemShut {NoStop}%
\bibitem [{\citenamefont {Cary}\ and\ \citenamefont
  {Littlejohn}(1983)}]{Cary_1983}%
  \BibitemOpen
  \bibfield  {author} {\bibinfo {author} {\bibfnamefont {J.~R.}\ \bibnamefont
  {Cary}}\ and\ \bibinfo {author} {\bibfnamefont {R.~G.}\ \bibnamefont
  {Littlejohn}},\ }\href@noop {} {\bibfield  {journal} {\bibinfo  {journal}
  {Annals of Physics}\ }\textbf {\bibinfo {volume} {151}},\ \bibinfo {pages}
  {1} (\bibinfo {year} {1983})}\BibitemShut {NoStop}%
\bibitem [{\citenamefont {Cary}\ and\ \citenamefont
  {Brizard}(2009)}]{Cary_2009}%
  \BibitemOpen
  \bibfield  {author} {\bibinfo {author} {\bibfnamefont {J.~R.}\ \bibnamefont
  {Cary}}\ and\ \bibinfo {author} {\bibfnamefont {A.~J.}\ \bibnamefont
  {Brizard}},\ }\href@noop {} {\bibfield  {journal} {\bibinfo  {journal}
  {Reviews of Modern Physics}\ }\textbf {\bibinfo {volume} {81}},\ \bibinfo
  {pages} {693} (\bibinfo {year} {2009})}\BibitemShut {NoStop}%
\bibitem [{\citenamefont {Landau}\ and\ \citenamefont
  {Lifshitz}(1969)}]{Landau_Lifshitz_1969}%
  \BibitemOpen
  \bibfield  {author} {\bibinfo {author} {\bibfnamefont {L.~D.}\ \bibnamefont
  {Landau}}\ and\ \bibinfo {author} {\bibfnamefont {E.~M.}\ \bibnamefont
  {Lifshitz}},\ }\href@noop {} {\emph {\bibinfo {title} {Mechanics}}}\
  (\bibinfo  {publisher} {Pergamon Press},\ \bibinfo {year} {1969})\BibitemShut
  {NoStop}%
\bibitem [{\citenamefont {Lichtenberg}\ and\ \citenamefont
  {Lieberman}(1983)}]{Lichtenberg_1983}%
  \BibitemOpen
  \bibfield  {author} {\bibinfo {author} {\bibfnamefont {A.~J.}\ \bibnamefont
  {Lichtenberg}}\ and\ \bibinfo {author} {\bibfnamefont {M.~A.}\ \bibnamefont
  {Lieberman}},\ }\href@noop {} {\emph {\bibinfo {title} {Regular and
  Stochastic Motion}}}\ (\bibinfo  {publisher} {Springer Science + Business
  Media},\ \bibinfo {year} {1983})\BibitemShut {NoStop}%
\bibitem [{\citenamefont {Marsden}\ and\ \citenamefont
  {Ratiu}(1999)}]{Marsden_1999}%
  \BibitemOpen
  \bibfield  {author} {\bibinfo {author} {\bibfnamefont {J.~E.}\ \bibnamefont
  {Marsden}}\ and\ \bibinfo {author} {\bibfnamefont {T.~S.}\ \bibnamefont
  {Ratiu}},\ }\href@noop {} {\emph {\bibinfo {title} {Introduction to Mechanics
  and Symmetry}}}\ (\bibinfo  {publisher} {Springer Science and Business
  Media},\ \bibinfo {year} {1999})\BibitemShut {NoStop}%
\bibitem [{\citenamefont {Hairer}, \citenamefont {Lubich},\ and\ \citenamefont
  {Wanner}(2006{\natexlab{a}})}]{Hairer_2006_symplectic}%
  \BibitemOpen
  \bibfield  {author} {\bibinfo {author} {\bibfnamefont {E.}~\bibnamefont
  {Hairer}}, \bibinfo {author} {\bibfnamefont {C.}~\bibnamefont {Lubich}}, \
  and\ \bibinfo {author} {\bibfnamefont {G.}~\bibnamefont {Wanner}},\ }\enquote
  {\bibinfo {title} {Geometric numerical integration},}\ \ (\bibinfo
  {publisher} {Springer},\ \bibinfo {year} {2006})\ pp.\ \bibinfo {pages}
  {179--236}\BibitemShut {NoStop}%
\bibitem [{\citenamefont {Mc{L}achlan}\ and\ \citenamefont
  {Quispel}(2006)}]{McLachlan_2006}%
  \BibitemOpen
  \bibfield  {author} {\bibinfo {author} {\bibfnamefont {R.~I.}\ \bibnamefont
  {Mc{L}achlan}}\ and\ \bibinfo {author} {\bibfnamefont {G.~R.~W.}\
  \bibnamefont {Quispel}},\ }\href@noop {} {\bibfield  {journal} {\bibinfo
  {journal} {Journal of Physics A: Mathematical and General}\ }\textbf
  {\bibinfo {volume} {39}},\ \bibinfo {pages} {5251} (\bibinfo {year}
  {2006})}\BibitemShut {NoStop}%
\bibitem [{\citenamefont {Forest}\ and\ \citenamefont
  {Ruth}(1990)}]{Forest_1990}%
  \BibitemOpen
  \bibfield  {author} {\bibinfo {author} {\bibfnamefont {E.}~\bibnamefont
  {Forest}}\ and\ \bibinfo {author} {\bibfnamefont {R.~D.}\ \bibnamefont
  {Ruth}},\ }\href@noop {} {\bibfield  {journal} {\bibinfo  {journal} {Physica
  D}\ }\textbf {\bibinfo {volume} {43}},\ \bibinfo {pages} {105} (\bibinfo
  {year} {1990})}\BibitemShut {NoStop}%
\bibitem [{\citenamefont {Qin}\ and\ \citenamefont {Guan}(2008)}]{Qin_2008}%
  \BibitemOpen
  \bibfield  {author} {\bibinfo {author} {\bibfnamefont {H.}~\bibnamefont
  {Qin}}\ and\ \bibinfo {author} {\bibfnamefont {X.}~\bibnamefont {Guan}},\
  }\href@noop {} {\bibfield  {journal} {\bibinfo  {journal} {Physical Review
  Letters}\ }\textbf {\bibinfo {volume} {100}},\ \bibinfo {pages} {035006}
  (\bibinfo {year} {2008})}\BibitemShut {NoStop}%
\bibitem [{\citenamefont {Qin}, \citenamefont {Guan},\ and\ \citenamefont
  {Tang}(2009)}]{Qin_2009}%
  \BibitemOpen
  \bibfield  {author} {\bibinfo {author} {\bibfnamefont {H.}~\bibnamefont
  {Qin}}, \bibinfo {author} {\bibfnamefont {X.}~\bibnamefont {Guan}}, \ and\
  \bibinfo {author} {\bibfnamefont {W.~M.}\ \bibnamefont {Tang}},\ }\href@noop
  {} {\bibfield  {journal} {\bibinfo  {journal} {Physics of Plasmas}\ }\textbf
  {\bibinfo {volume} {16}},\ \bibinfo {pages} {042510} (\bibinfo {year}
  {2009})}\BibitemShut {NoStop}%
\bibitem [{\citenamefont {Squire}, \citenamefont {Qin},\ and\ \citenamefont
  {Tang}(2012{\natexlab{a}})}]{Squire_2012_PIC}%
  \BibitemOpen
  \bibfield  {author} {\bibinfo {author} {\bibfnamefont {J.}~\bibnamefont
  {Squire}}, \bibinfo {author} {\bibfnamefont {H.}~\bibnamefont {Qin}}, \ and\
  \bibinfo {author} {\bibfnamefont {W.~M.}\ \bibnamefont {Tang}},\ }\href@noop
  {} {\bibfield  {journal} {\bibinfo  {journal} {Physics of Plasmas}\ }\textbf
  {\bibinfo {volume} {19}},\ \bibinfo {pages} {084501} (\bibinfo {year}
  {2012}{\natexlab{a}})}\BibitemShut {NoStop}%
\bibitem [{\citenamefont {Qin}\ \emph {et~al.}(2016)\citenamefont {Qin},
  \citenamefont {Liu}, \citenamefont {Xiao}, \citenamefont {Zhang},
  \citenamefont {He}, \citenamefont {Sun}, \citenamefont {Burby}, \citenamefont
  {Ellison},\ and\ \citenamefont {Zhou}}]{Qin_2016}%
  \BibitemOpen
  \bibfield  {author} {\bibinfo {author} {\bibfnamefont {H.}~\bibnamefont
  {Qin}}, \bibinfo {author} {\bibfnamefont {J.}~\bibnamefont {Liu}}, \bibinfo
  {author} {\bibfnamefont {J.}~\bibnamefont {Xiao}}, \bibinfo {author}
  {\bibfnamefont {R.}~\bibnamefont {Zhang}}, \bibinfo {author} {\bibfnamefont
  {Y.}~\bibnamefont {He}}, \bibinfo {author} {\bibfnamefont {Y.}~\bibnamefont
  {Sun}}, \bibinfo {author} {\bibfnamefont {J.}~\bibnamefont {Burby}}, \bibinfo
  {author} {\bibfnamefont {C.}~\bibnamefont {Ellison}}, \ and\ \bibinfo
  {author} {\bibfnamefont {Y.}~\bibnamefont {Zhou}},\ }\href@noop {} {\bibfield
   {journal} {\bibinfo  {journal} {Nuclear Fusion}\ }\textbf {\bibinfo {volume}
  {56}},\ \bibinfo {pages} {014001} (\bibinfo {year} {2016})}\BibitemShut
  {NoStop}%
\bibitem [{\citenamefont {Darboux}(1882)}]{Darboux_1882}%
  \BibitemOpen
  \bibfield  {author} {\bibinfo {author} {\bibfnamefont {G.}~\bibnamefont
  {Darboux}},\ }\href@noop {} {\bibfield  {journal} {\bibinfo  {journal}
  {Bulletin des Sciences Math{\'e}matiques}\ }\textbf {\bibinfo {volume} {6}},\
  \bibinfo {pages} {14} (\bibinfo {year} {1882})}\BibitemShut {NoStop}%
\bibitem [{\citenamefont {Karas{\"o}zen}(2004)}]{Karasozen_2004}%
  \BibitemOpen
  \bibfield  {author} {\bibinfo {author} {\bibfnamefont {B.}~\bibnamefont
  {Karas{\"o}zen}},\ }\href@noop {} {\bibfield  {journal} {\bibinfo  {journal}
  {Mathematical and Computer Modelling}\ }\textbf {\bibinfo {volume} {40}},\
  \bibinfo {pages} {1225} (\bibinfo {year} {2004})}\BibitemShut {NoStop}%
\bibitem [{\citenamefont {White}\ and\ \citenamefont
  {Chance}(1984)}]{White_1984}%
  \BibitemOpen
  \bibfield  {author} {\bibinfo {author} {\bibfnamefont {R.~B.}\ \bibnamefont
  {White}}\ and\ \bibinfo {author} {\bibfnamefont {M.~S.}\ \bibnamefont
  {Chance}},\ }\href@noop {} {\bibfield  {journal} {\bibinfo  {journal}
  {Physics of Fluids}\ }\textbf {\bibinfo {volume} {27}},\ \bibinfo {pages}
  {2455} (\bibinfo {year} {1984})}\BibitemShut {NoStop}%
\bibitem [{\citenamefont {Velasco}\ \emph {et~al.}(2012)\citenamefont
  {Velasco}, \citenamefont {Bustos}, \citenamefont {Castejon}, \citenamefont
  {Fernandez}, \citenamefont {Martin-Mayor},\ and\ \citenamefont
  {Tarancon}}]{Velasco_2012}%
  \BibitemOpen
  \bibfield  {author} {\bibinfo {author} {\bibfnamefont {J.~L.}\ \bibnamefont
  {Velasco}}, \bibinfo {author} {\bibfnamefont {A.}~\bibnamefont {Bustos}},
  \bibinfo {author} {\bibfnamefont {F.}~\bibnamefont {Castejon}}, \bibinfo
  {author} {\bibfnamefont {L.~A.}\ \bibnamefont {Fernandez}}, \bibinfo {author}
  {\bibfnamefont {V.}~\bibnamefont {Martin-Mayor}}, \ and\ \bibinfo {author}
  {\bibfnamefont {A.}~\bibnamefont {Tarancon}},\ }\href@noop {} {\bibfield
  {journal} {\bibinfo  {journal} {Computer Physics Communications}\ }\textbf
  {\bibinfo {volume} {183}},\ \bibinfo {pages} {1877} (\bibinfo {year}
  {2012})}\BibitemShut {NoStop}%
\bibitem [{\citenamefont {Kramer}\ \emph {et~al.}(2013)\citenamefont {Kramer},
  \citenamefont {Budny}, \citenamefont {Bortolon}, \citenamefont {Fredrickson},
  \citenamefont {Fu}, \citenamefont {Heidbrink}, \citenamefont {Nazikian},
  \citenamefont {Valeo},\ and\ \citenamefont {Zeeland}}]{Kramer_2013}%
  \BibitemOpen
  \bibfield  {author} {\bibinfo {author} {\bibfnamefont {G.~J.}\ \bibnamefont
  {Kramer}}, \bibinfo {author} {\bibfnamefont {R.~V.}\ \bibnamefont {Budny}},
  \bibinfo {author} {\bibfnamefont {A.}~\bibnamefont {Bortolon}}, \bibinfo
  {author} {\bibfnamefont {E.~D.}\ \bibnamefont {Fredrickson}}, \bibinfo
  {author} {\bibfnamefont {G.~Y.}\ \bibnamefont {Fu}}, \bibinfo {author}
  {\bibfnamefont {W.~W.}\ \bibnamefont {Heidbrink}}, \bibinfo {author}
  {\bibfnamefont {R.}~\bibnamefont {Nazikian}}, \bibinfo {author}
  {\bibfnamefont {E.}~\bibnamefont {Valeo}}, \ and\ \bibinfo {author}
  {\bibfnamefont {M.~A.~V.}\ \bibnamefont {Zeeland}},\ }\href {\doibase
  10.1088/0741-3335/55/2/025013} {\bibfield  {journal} {\bibinfo  {journal}
  {Plasma Physics and Controlled Fusion}\ }\textbf {\bibinfo {volume} {55}},\
  \bibinfo {pages} {025013} (\bibinfo {year} {2013})}\BibitemShut {NoStop}%
\bibitem [{\citenamefont {Pfefferl{\'e}}\ \emph {et~al.}(2014)\citenamefont
  {Pfefferl{\'e}}, \citenamefont {Graves}, \citenamefont {Cooper},
  \citenamefont {Misev}, \citenamefont {Chapman},\ and\ \citenamefont {ans
  S~Sangaroon}}]{Pfefferle_2014}%
  \BibitemOpen
  \bibfield  {author} {\bibinfo {author} {\bibfnamefont {D.}~\bibnamefont
  {Pfefferl{\'e}}}, \bibinfo {author} {\bibfnamefont {J.~P.}\ \bibnamefont
  {Graves}}, \bibinfo {author} {\bibfnamefont {W.~A.}\ \bibnamefont {Cooper}},
  \bibinfo {author} {\bibfnamefont {C.}~\bibnamefont {Misev}}, \bibinfo
  {author} {\bibfnamefont {I.~T.}\ \bibnamefont {Chapman}}, \ and\ \bibinfo
  {author} {\bibfnamefont {M.~T.}\ \bibnamefont {ans S~Sangaroon}},\
  }\href@noop {} {\bibfield  {journal} {\bibinfo  {journal} {Nuclear Fusion}\
  }\textbf {\bibinfo {volume} {54}},\ \bibinfo {pages} {064020} (\bibinfo
  {year} {2014})}\BibitemShut {NoStop}%
\bibitem [{\citenamefont {Hirvijoki}\ \emph {et~al.}(2014)\citenamefont
  {Hirvijoki}, \citenamefont {Asunta}, \citenamefont {Koskela}, \citenamefont
  {Kurki-Suonio}, \citenamefont {Meittunen}, \citenamefont {Sipil{\"a}},
  \citenamefont {Snicker},\ and\ \citenamefont
  {{\"A}k{\"a}slompolo}}]{Hirvijoki_2014}%
  \BibitemOpen
  \bibfield  {author} {\bibinfo {author} {\bibfnamefont {E.}~\bibnamefont
  {Hirvijoki}}, \bibinfo {author} {\bibfnamefont {O.}~\bibnamefont {Asunta}},
  \bibinfo {author} {\bibfnamefont {T.}~\bibnamefont {Koskela}}, \bibinfo
  {author} {\bibfnamefont {T.}~\bibnamefont {Kurki-Suonio}}, \bibinfo {author}
  {\bibfnamefont {J.}~\bibnamefont {Meittunen}}, \bibinfo {author}
  {\bibfnamefont {S.}~\bibnamefont {Sipil{\"a}}}, \bibinfo {author}
  {\bibfnamefont {A.}~\bibnamefont {Snicker}}, \ and\ \bibinfo {author}
  {\bibfnamefont {S.}~\bibnamefont {{\"A}k{\"a}slompolo}},\ }\href@noop {}
  {\bibfield  {journal} {\bibinfo  {journal} {Computational Physics
  Communications}\ }\textbf {\bibinfo {volume} {185}},\ \bibinfo {pages} {1310}
  (\bibinfo {year} {2014})}\BibitemShut {NoStop}%
\bibitem [{\citenamefont {Hirvijoki}\ \emph {et~al.}(2015)\citenamefont
  {Hirvijoki}, \citenamefont {Kurki-Suonio}, \citenamefont
  {{\"A}k{\"a}slompolo}, \citenamefont {Varje}, \citenamefont {Koskela},\ and\
  \citenamefont {Meittunen}}]{Hirvijoki_2015}%
  \BibitemOpen
  \bibfield  {author} {\bibinfo {author} {\bibfnamefont {E.}~\bibnamefont
  {Hirvijoki}}, \bibinfo {author} {\bibfnamefont {T.}~\bibnamefont
  {Kurki-Suonio}}, \bibinfo {author} {\bibfnamefont {S.}~\bibnamefont
  {{\"A}k{\"a}slompolo}}, \bibinfo {author} {\bibfnamefont {J.}~\bibnamefont
  {Varje}}, \bibinfo {author} {\bibfnamefont {T.}~\bibnamefont {Koskela}}, \
  and\ \bibinfo {author} {\bibfnamefont {J.}~\bibnamefont {Meittunen}},\
  }\href@noop {} {\bibfield  {journal} {\bibinfo  {journal} {Journal of Plasma
  Physics}\ }\textbf {\bibinfo {volume} {81}},\ \bibinfo {pages} {435810301}
  (\bibinfo {year} {2015})}\BibitemShut {NoStop}%
\bibitem [{\citenamefont {Li}\ \emph {et~al.}(2011)\citenamefont {Li},
  \citenamefont {Qin}, \citenamefont {Pu}, \citenamefont {Xie},\ and\
  \citenamefont {Fu}}]{Li_2011}%
  \BibitemOpen
  \bibfield  {author} {\bibinfo {author} {\bibfnamefont {J.}~\bibnamefont
  {Li}}, \bibinfo {author} {\bibfnamefont {H.}~\bibnamefont {Qin}}, \bibinfo
  {author} {\bibfnamefont {Z.}~\bibnamefont {Pu}}, \bibinfo {author}
  {\bibfnamefont {L.}~\bibnamefont {Xie}}, \ and\ \bibinfo {author}
  {\bibfnamefont {S.}~\bibnamefont {Fu}},\ }\href@noop {} {\bibfield  {journal}
  {\bibinfo  {journal} {Physics of Plasmas}\ }\textbf {\bibinfo {volume}
  {18}},\ \bibinfo {pages} {052902} (\bibinfo {year} {2011})}\BibitemShut
  {NoStop}%
\bibitem [{\citenamefont {Kraus}(2013)}]{Kraus_2013}%
  \BibitemOpen
  \bibfield  {author} {\bibinfo {author} {\bibfnamefont {M.}~\bibnamefont
  {Kraus}},\ }\emph {\bibinfo {title} {Variational Integrators in Plasma
  Physics}},\ \href@noop {} {\bibinfo {type} {Doctoral {T}hesis}},\ \bibinfo
  {school} {Technische Universit{\"a}t M{\"u}nchen} (\bibinfo {year}
  {2013})\BibitemShut {NoStop}%
\bibitem [{\citenamefont {Marsden}\ and\ \citenamefont
  {West}(2001)}]{Marsden_2001}%
  \BibitemOpen
  \bibfield  {author} {\bibinfo {author} {\bibfnamefont {J.~E.}\ \bibnamefont
  {Marsden}}\ and\ \bibinfo {author} {\bibfnamefont {M.}~\bibnamefont {West}},\
  }\href {\doibase 10.1017/S096249290100006X} {\bibfield  {journal} {\bibinfo
  {journal} {Acta Numerica}\ }\textbf {\bibinfo {volume} {10}},\ \bibinfo
  {pages} {357–514} (\bibinfo {year} {2001})}\BibitemShut {NoStop}%
\bibitem [{\citenamefont {Squire}, \citenamefont {Qin},\ and\ \citenamefont
  {Tang}(2012{\natexlab{b}})}]{Squire_2012}%
  \BibitemOpen
  \bibfield  {author} {\bibinfo {author} {\bibfnamefont {J.}~\bibnamefont
  {Squire}}, \bibinfo {author} {\bibfnamefont {H.}~\bibnamefont {Qin}}, \ and\
  \bibinfo {author} {\bibfnamefont {W.~M.}\ \bibnamefont {Tang}},\ }\href@noop
  {} {\bibfield  {journal} {\bibinfo  {journal} {Physics of Plasmas}\ }\textbf
  {\bibinfo {volume} {19}},\ \bibinfo {pages} {052501} (\bibinfo {year}
  {2012}{\natexlab{b}})}\BibitemShut {NoStop}%
\bibitem [{\citenamefont {Ellison}\ \emph {et~al.}(2015)\citenamefont
  {Ellison}, \citenamefont {Finn}, \citenamefont {Qin},\ and\ \citenamefont
  {Tang}}]{Ellison_2015_PPCF}%
  \BibitemOpen
  \bibfield  {author} {\bibinfo {author} {\bibfnamefont {C.~L.}\ \bibnamefont
  {Ellison}}, \bibinfo {author} {\bibfnamefont {J.~M.}\ \bibnamefont {Finn}},
  \bibinfo {author} {\bibfnamefont {H.}~\bibnamefont {Qin}}, \ and\ \bibinfo
  {author} {\bibfnamefont {W.~M.}\ \bibnamefont {Tang}},\ }\href {\doibase
  10.1088/0741-3335/57/5/054007} {\bibfield  {journal} {\bibinfo  {journal}
  {Plasma Physics and Controlled Fusion}\ }\textbf {\bibinfo {volume} {57}},\
  \bibinfo {pages} {054007} (\bibinfo {year} {2015})}\BibitemShut {NoStop}%
\bibitem [{\citenamefont {Ellison}(2016)}]{Ellison_thesis}%
  \BibitemOpen
  \bibfield  {author} {\bibinfo {author} {\bibfnamefont {C.~L.}\ \bibnamefont
  {Ellison}},\ }\emph {\bibinfo {title} {Development of Multistep and
  Degenerate Variational Integrators for Applications in Plasma Physics}},\
  \href@noop {} {\bibinfo {type} {Doctoral {T}hesis}},\ \bibinfo  {school}
  {Princeton University} (\bibinfo {year} {2016})\BibitemShut {NoStop}%
\bibitem [{\citenamefont {Goldstein}, \citenamefont {Poole},\ and\
  \citenamefont {Safko}(2001{\natexlab{a}})}]{Goldstein_2001}%
  \BibitemOpen
  \bibfield  {author} {\bibinfo {author} {\bibfnamefont {H.}~\bibnamefont
  {Goldstein}}, \bibinfo {author} {\bibfnamefont {C.}~\bibnamefont {Poole}}, \
  and\ \bibinfo {author} {\bibfnamefont {J.}~\bibnamefont {Safko}},\
  }\href@noop {} {\emph {\bibinfo {title} {Classical Mechanics}}}\ (\bibinfo
  {publisher} {Addison Wesley},\ \bibinfo {year} {2001})\BibitemShut {NoStop}%
\bibitem [{\citenamefont {Rowley}\ and\ \citenamefont
  {Marsden}(2002)}]{Rowley_2002}%
  \BibitemOpen
  \bibfield  {author} {\bibinfo {author} {\bibfnamefont {C.~W.}\ \bibnamefont
  {Rowley}}\ and\ \bibinfo {author} {\bibfnamefont {J.~E.}\ \bibnamefont
  {Marsden}},\ }\href@noop {} {\bibfield  {journal} {\bibinfo  {journal}
  {Proceedings of the 41st IEEE Conference on Decision and Control}\ }\textbf
  {\bibinfo {volume} {2}},\ \bibinfo {pages} {1521} (\bibinfo {year}
  {2002})}\BibitemShut {NoStop}%
\bibitem [{\citenamefont {Ober-Bl{\"o}baum}\ \emph {et~al.}(2013)\citenamefont
  {Ober-Bl{\"o}baum}, \citenamefont {Tao}, \citenamefont {Cheng}, \citenamefont
  {Owhadi},\ and\ \citenamefont {Marsden}}]{Ober-Blobaum_2013}%
  \BibitemOpen
  \bibfield  {author} {\bibinfo {author} {\bibfnamefont {S.}~\bibnamefont
  {Ober-Bl{\"o}baum}}, \bibinfo {author} {\bibfnamefont {M.}~\bibnamefont
  {Tao}}, \bibinfo {author} {\bibfnamefont {M.}~\bibnamefont {Cheng}}, \bibinfo
  {author} {\bibfnamefont {H.}~\bibnamefont {Owhadi}}, \ and\ \bibinfo {author}
  {\bibfnamefont {J.~E.}\ \bibnamefont {Marsden}},\ }\href@noop {} {\bibfield
  {journal} {\bibinfo  {journal} {Journal of Computational Physics}\ }
  (\bibinfo {year} {2013})}\BibitemShut {NoStop}%
\bibitem [{\citenamefont {Tyranowski}\ and\ \citenamefont
  {Desbrun}(2014)}]{Tyranowski_2014}%
  \BibitemOpen
  \bibfield  {author} {\bibinfo {author} {\bibfnamefont {T.}~\bibnamefont
  {Tyranowski}}\ and\ \bibinfo {author} {\bibfnamefont {M.}~\bibnamefont
  {Desbrun}},\ }\href@noop {} {\bibfield  {journal} {\bibinfo  {journal}
  {arXiv:1401.7904}\ } (\bibinfo {year} {2014})}\BibitemShut {NoStop}%
\bibitem [{\citenamefont {Burby}\ and\ \citenamefont
  {Ellison}(2017)}]{Burby_2017}%
  \BibitemOpen
  \bibfield  {author} {\bibinfo {author} {\bibfnamefont {J.~W.}\ \bibnamefont
  {Burby}}\ and\ \bibinfo {author} {\bibfnamefont {C.~L.}\ \bibnamefont
  {Ellison}},\ }\href {\doibase 10.1063/1.5004429} {\bibfield  {journal}
  {\bibinfo  {journal} {Physics of Plasmas}\ }\textbf {\bibinfo {volume}
  {24}},\ \bibinfo {pages} {110703} (\bibinfo {year} {2017})},\ \Eprint
  {http://arxiv.org/abs/https://doi.org/10.1063/1.5004429}
  {https://doi.org/10.1063/1.5004429} \BibitemShut {NoStop}%
\bibitem [{\citenamefont {Zhang}, \citenamefont {Jia},\ and\ \citenamefont
  {Sun}(2014)}]{Zhang_2014}%
  \BibitemOpen
  \bibfield  {author} {\bibinfo {author} {\bibfnamefont {S.}~\bibnamefont
  {Zhang}}, \bibinfo {author} {\bibfnamefont {Y.}~\bibnamefont {Jia}}, \ and\
  \bibinfo {author} {\bibfnamefont {Q.}~\bibnamefont {Sun}},\ }\href@noop {}
  {\bibfield  {journal} {\bibinfo  {journal} {Journal of Computational
  Physics}\ }\textbf {\bibinfo {volume} {282}},\ \bibinfo {pages} {43}
  (\bibinfo {year} {2014})}\BibitemShut {NoStop}%
\bibitem [{\citenamefont {Kraus}(2017)}]{Kraus_2017_PVI}%
  \BibitemOpen
  \bibfield  {author} {\bibinfo {author} {\bibfnamefont {M.}~\bibnamefont
  {Kraus}},\ }\href@noop {} {\enquote {\bibinfo {title} {Projected variational
  integrators for degenerate lagrangian systems},}\ } (\bibinfo {year}
  {2017}),\ \Eprint {http://arxiv.org/abs/1708.07356} {arXiv:1708.07356
  [math.NA]} \BibitemShut {NoStop}%
\bibitem [{\citenamefont {Ruth}(1983)}]{Ruth_1983}%
  \BibitemOpen
  \bibfield  {author} {\bibinfo {author} {\bibfnamefont {R.~D.}\ \bibnamefont
  {Ruth}},\ }\href@noop {} {\bibfield  {journal} {\bibinfo  {journal} {IEEE
  Transactions on Nuclear Science}\ }\textbf {\bibinfo {volume} {NS-30}},\
  \bibinfo {pages} {2669} (\bibinfo {year} {1983})}\BibitemShut {NoStop}%
\bibitem [{\citenamefont {Channell}\ and\ \citenamefont
  {Scovel}(1990)}]{Channell_1990}%
  \BibitemOpen
  \bibfield  {author} {\bibinfo {author} {\bibfnamefont {P.~J.}\ \bibnamefont
  {Channell}}\ and\ \bibinfo {author} {\bibfnamefont {C.}~\bibnamefont
  {Scovel}},\ }\href@noop {} {\bibfield  {journal} {\bibinfo  {journal}
  {Nonlinearity}\ }\textbf {\bibinfo {volume} {3}},\ \bibinfo {pages} {231}
  (\bibinfo {year} {1990})}\BibitemShut {NoStop}%
\bibitem [{\citenamefont {Sanz-Serna}(1988)}]{Sanz-Serna_1988}%
  \BibitemOpen
  \bibfield  {author} {\bibinfo {author} {\bibfnamefont {J.~M.}\ \bibnamefont
  {Sanz-Serna}},\ }\href@noop {} {\bibfield  {journal} {\bibinfo  {journal}
  {BIT}\ }\textbf {\bibinfo {volume} {28}},\ \bibinfo {pages} {877} (\bibinfo
  {year} {1988})}\BibitemShut {NoStop}%
\bibitem [{\citenamefont {Leok}\ and\ \citenamefont {Zhang}(2011)}]{Leok_2011}%
  \BibitemOpen
  \bibfield  {author} {\bibinfo {author} {\bibfnamefont {M.}~\bibnamefont
  {Leok}}\ and\ \bibinfo {author} {\bibfnamefont {J.}~\bibnamefont {Zhang}},\
  }\href@noop {} {\bibfield  {journal} {\bibinfo  {journal} {IMA Journal of
  Numerical Analysis}\ }\textbf {\bibinfo {volume} {31}},\ \bibinfo {pages}
  {1497} (\bibinfo {year} {2011})}\BibitemShut {NoStop}%
\bibitem [{\citenamefont {Arnold}(1989)}]{Arnold_1989_phase_space_action}%
  \BibitemOpen
  \bibfield  {author} {\bibinfo {author} {\bibfnamefont {V.~I.}\ \bibnamefont
  {Arnold}},\ }\href@noop {} {\emph {\bibinfo {title} {Mathematical Methods of
  Classical Mechanics}}}\ (\bibinfo  {publisher} {Springer},\ \bibinfo {year}
  {1989})\ p.\ \bibinfo {pages} {243}\BibitemShut {NoStop}%
\bibitem [{\citenamefont {Goldstein}, \citenamefont {Poole},\ and\
  \citenamefont
  {Safko}(2001{\natexlab{b}})}]{Goldstein_2001_phase_space_action}%
  \BibitemOpen
  \bibfield  {author} {\bibinfo {author} {\bibfnamefont {H.}~\bibnamefont
  {Goldstein}}, \bibinfo {author} {\bibfnamefont {C.}~\bibnamefont {Poole}}, \
  and\ \bibinfo {author} {\bibfnamefont {J.}~\bibnamefont {Safko}},\
  }\href@noop {} {\emph {\bibinfo {title} {Classical Mechanics}}}\ (\bibinfo
  {publisher} {Addison Wesley},\ \bibinfo {year} {2001})\ Chap.\ \bibinfo
  {chapter} {8.5}, p.\ \bibinfo {pages} {353}\BibitemShut {NoStop}%
\bibitem [{\citenamefont {Holm}, \citenamefont {Schmah},\ and\ \citenamefont
  {Stoica}(2009)}]{Holm_2009}%
  \BibitemOpen
  \bibfield  {author} {\bibinfo {author} {\bibfnamefont {D.~D.}\ \bibnamefont
  {Holm}}, \bibinfo {author} {\bibfnamefont {T.}~\bibnamefont {Schmah}}, \ and\
  \bibinfo {author} {\bibfnamefont {C.}~\bibnamefont {Stoica}},\ }\href@noop {}
  {\emph {\bibinfo {title} {Geometric Mechanics and Symmetry: From Finite to
  Infinite Dimensions}}}\ (\bibinfo  {publisher} {Oxford University Press},\
  \bibinfo {year} {2009})\BibitemShut {NoStop}%
\bibitem [{\citenamefont {Dahlquist}(1956)}]{Dahlquist_1956}%
  \BibitemOpen
  \bibfield  {author} {\bibinfo {author} {\bibfnamefont {G.}~\bibnamefont
  {Dahlquist}},\ }\href@noop {} {\bibfield  {journal} {\bibinfo  {journal}
  {Mathematica Scandinavica}\ }\textbf {\bibinfo {volume} {4}},\ \bibinfo
  {pages} {33} (\bibinfo {year} {1956})}\BibitemShut {NoStop}%
\bibitem [{\citenamefont {Hairer}, \citenamefont {Lubich},\ and\ \citenamefont
  {Wanner}(2006{\natexlab{b}})}]{Hairer_2006_multistep}%
  \BibitemOpen
  \bibfield  {author} {\bibinfo {author} {\bibfnamefont {E.}~\bibnamefont
  {Hairer}}, \bibinfo {author} {\bibfnamefont {C.}~\bibnamefont {Lubich}}, \
  and\ \bibinfo {author} {\bibfnamefont {G.}~\bibnamefont {Wanner}},\ }\enquote
  {\bibinfo {title} {Geometric numerical integration},}\ \ (\bibinfo
  {publisher} {Springer},\ \bibinfo {year} {2006})\ pp.\ \bibinfo {pages}
  {567--616}\BibitemShut {NoStop}%
\bibitem [{\citenamefont {Hairer}(1999)}]{Hairer_1999}%
  \BibitemOpen
  \bibfield  {author} {\bibinfo {author} {\bibfnamefont {E.}~\bibnamefont
  {Hairer}},\ }\href@noop {} {\bibfield  {journal} {\bibinfo  {journal}
  {Numerische Mathematik}\ }\textbf {\bibinfo {volume} {84}},\ \bibinfo {pages}
  {199} (\bibinfo {year} {1999})}\BibitemShut {NoStop}%
\bibitem [{\citenamefont {Bashforth}\ and\ \citenamefont
  {Adams}(1883)}]{Bashforth_1883}%
  \BibitemOpen
  \bibfield  {author} {\bibinfo {author} {\bibfnamefont {F.}~\bibnamefont
  {Bashforth}}\ and\ \bibinfo {author} {\bibfnamefont {J.~C.}\ \bibnamefont
  {Adams}},\ }\href@noop {} {\emph {\bibinfo {title} {An attempt to test the
  theories of capillary action by comparing the theoretical and measured forms
  of drops of fluid, with an explanation of the method of integration employed
  in constructing the tables which give the theoretical forms of such drops}}}\
  (\bibinfo  {publisher} {Cambridge University Press},\ \bibinfo {year}
  {1883})\BibitemShut {NoStop}%
\bibitem [{\citenamefont {Hairer}, \citenamefont {Lubich},\ and\ \citenamefont
  {Wanner}(2006{\natexlab{c}})}]{Hairer_2006_seuler}%
  \BibitemOpen
  \bibfield  {author} {\bibinfo {author} {\bibfnamefont {E.}~\bibnamefont
  {Hairer}}, \bibinfo {author} {\bibfnamefont {C.}~\bibnamefont {Lubich}}, \
  and\ \bibinfo {author} {\bibfnamefont {G.}~\bibnamefont {Wanner}},\ }\enquote
  {\bibinfo {title} {Geometric numerical integration},}\ \ (\bibinfo
  {publisher} {Springer},\ \bibinfo {year} {2006})\ pp.\ \bibinfo {pages}
  {3--4}\BibitemShut {NoStop}%
\bibitem [{\citenamefont {Hairer}, \citenamefont {Lubich},\ and\ \citenamefont
  {Wanner}(2006{\natexlab{d}})}]{Hairer_2006_adjoint}%
  \BibitemOpen
  \bibfield  {author} {\bibinfo {author} {\bibfnamefont {E.}~\bibnamefont
  {Hairer}}, \bibinfo {author} {\bibfnamefont {C.}~\bibnamefont {Lubich}}, \
  and\ \bibinfo {author} {\bibfnamefont {G.}~\bibnamefont {Wanner}},\ }\enquote
  {\bibinfo {title} {Geometric numerical integration},}\ \ (\bibinfo
  {publisher} {Springer},\ \bibinfo {year} {2006})\ p.~\bibinfo {pages}
  {42}\BibitemShut {NoStop}%
\bibitem [{\citenamefont {Blanes}, \citenamefont {Casas},\ and\ \citenamefont
  {Murua}(2004)}]{Blanes_2004}%
  \BibitemOpen
  \bibfield  {author} {\bibinfo {author} {\bibfnamefont {S.}~\bibnamefont
  {Blanes}}, \bibinfo {author} {\bibfnamefont {F.}~\bibnamefont {Casas}}, \
  and\ \bibinfo {author} {\bibfnamefont {A.}~\bibnamefont {Murua}},\ }\href
  {\doibase 10.1137/S0036142902417029} {\bibfield  {journal} {\bibinfo
  {journal} {SIAM Journal on Numerical Analysis}\ }\textbf {\bibinfo {volume}
  {42}},\ \bibinfo {pages} {531} (\bibinfo {year} {2004})}\BibitemShut
  {NoStop}%
\bibitem [{\citenamefont {He}\ \emph {et~al.}(2016)\citenamefont {He},
  \citenamefont {Sun}, \citenamefont {Zhang}, \citenamefont {Wang},
  \citenamefont {Liu},\ and\ \citenamefont {Qin}}]{He_2016}%
  \BibitemOpen
  \bibfield  {author} {\bibinfo {author} {\bibfnamefont {Y.}~\bibnamefont
  {He}}, \bibinfo {author} {\bibfnamefont {Y.}~\bibnamefont {Sun}}, \bibinfo
  {author} {\bibfnamefont {R.}~\bibnamefont {Zhang}}, \bibinfo {author}
  {\bibfnamefont {Y.}~\bibnamefont {Wang}}, \bibinfo {author} {\bibfnamefont
  {J.}~\bibnamefont {Liu}}, \ and\ \bibinfo {author} {\bibfnamefont
  {H.}~\bibnamefont {Qin}},\ }\href {\doibase 10.1063/1.4962677} {\bibfield
  {journal} {\bibinfo  {journal} {Physics of Plasmas}\ }\textbf {\bibinfo
  {volume} {23}},\ \bibinfo {pages} {092109} (\bibinfo {year}
  {2016})}\BibitemShut {NoStop}%
\bibitem [{\citenamefont {Northrop}(1963)}]{Northrop_1963}%
  \BibitemOpen
  \bibfield  {author} {\bibinfo {author} {\bibfnamefont {T.~G.}\ \bibnamefont
  {Northrop}},\ }\href {\doibase 10.1029/RG001i003p00283} {\bibfield  {journal}
  {\bibinfo  {journal} {Reviews of Geophysics}\ }\textbf {\bibinfo {volume}
  {1}},\ \bibinfo {pages} {283} (\bibinfo {year} {1963})}\BibitemShut {NoStop}%
\bibitem [{\citenamefont {White}\ and\ \citenamefont
  {Zakharov}(2003)}]{White_2003}%
  \BibitemOpen
  \bibfield  {author} {\bibinfo {author} {\bibfnamefont {R.}~\bibnamefont
  {White}}\ and\ \bibinfo {author} {\bibfnamefont {L.~E.}\ \bibnamefont
  {Zakharov}},\ }\href@noop {} {\bibfield  {journal} {\bibinfo  {journal}
  {Physics of Plasmas}\ }\textbf {\bibinfo {volume} {10}},\ \bibinfo {pages}
  {573} (\bibinfo {year} {2003})}\BibitemShut {NoStop}%
\bibitem [{\citenamefont {Yoshida}(1990)}]{Yoshida_1990}%
  \BibitemOpen
  \bibfield  {author} {\bibinfo {author} {\bibfnamefont {H.}~\bibnamefont
  {Yoshida}},\ }\href@noop {} {\bibfield  {journal} {\bibinfo  {journal}
  {Physics Letters A}\ }\textbf {\bibinfo {volume} {150}},\ \bibinfo {pages}
  {262} (\bibinfo {year} {1990})}\BibitemShut {NoStop}%
\bibitem [{\citenamefont {Friedman}\ and\ \citenamefont
  {Auerbach}(1991)}]{Friedman_1991}%
  \BibitemOpen
  \bibfield  {author} {\bibinfo {author} {\bibfnamefont {A.}~\bibnamefont
  {Friedman}}\ and\ \bibinfo {author} {\bibfnamefont {S.~P.}\ \bibnamefont
  {Auerbach}},\ }\href {\doibase https://doi.org/10.1016/0021-9991(91)90078-Y}
  {\bibfield  {journal} {\bibinfo  {journal} {Journal of Computational
  Physics}\ }\textbf {\bibinfo {volume} {93}},\ \bibinfo {pages} {171 }
  (\bibinfo {year} {1991})}\BibitemShut {NoStop}%
\bibitem [{\citenamefont {He}\ \emph {et~al.}(2015)\citenamefont {He},
  \citenamefont {Sun}, \citenamefont {Liu},\ and\ \citenamefont
  {Qin}}]{He_2015}%
  \BibitemOpen
  \bibfield  {author} {\bibinfo {author} {\bibfnamefont {Y.}~\bibnamefont
  {He}}, \bibinfo {author} {\bibfnamefont {Y.}~\bibnamefont {Sun}}, \bibinfo
  {author} {\bibfnamefont {J.}~\bibnamefont {Liu}}, \ and\ \bibinfo {author}
  {\bibfnamefont {H.}~\bibnamefont {Qin}},\ }\href@noop {} {\bibfield
  {journal} {\bibinfo  {journal} {Journal of Computational Physics}\ }\textbf
  {\bibinfo {volume} {281}},\ \bibinfo {pages} {135} (\bibinfo {year}
  {2015})}\BibitemShut {NoStop}%
\bibitem [{\citenamefont {Wang}\ \emph {et~al.}(2017)\citenamefont {Wang},
  \citenamefont {Liu}, \citenamefont {Qin}, \citenamefont {Yu},\ and\
  \citenamefont {Yao}}]{Wang_2017}%
  \BibitemOpen
  \bibfield  {author} {\bibinfo {author} {\bibfnamefont {Y.}~\bibnamefont
  {Wang}}, \bibinfo {author} {\bibfnamefont {J.}~\bibnamefont {Liu}}, \bibinfo
  {author} {\bibfnamefont {H.}~\bibnamefont {Qin}}, \bibinfo {author}
  {\bibfnamefont {Z.}~\bibnamefont {Yu}}, \ and\ \bibinfo {author}
  {\bibfnamefont {Y.}~\bibnamefont {Yao}},\ }\href {\doibase
  https://doi.org/10.1016/j.cpc.2017.07.009} {\bibfield  {journal} {\bibinfo
  {journal} {Computer Physics Communications}\ }\textbf {\bibinfo {volume}
  {220}},\ \bibinfo {pages} {212 } (\bibinfo {year} {2017})}\BibitemShut
  {NoStop}%
\bibitem [{\citenamefont {Kraus}\ \emph {et~al.}(2017)\citenamefont {Kraus},
  \citenamefont {Kormann}, \citenamefont {Morrison},\ and\ \citenamefont
  {Sonnendrücker}}]{Kraus_2017}%
  \BibitemOpen
  \bibfield  {author} {\bibinfo {author} {\bibfnamefont {M.}~\bibnamefont
  {Kraus}}, \bibinfo {author} {\bibfnamefont {K.}~\bibnamefont {Kormann}},
  \bibinfo {author} {\bibfnamefont {P.}~\bibnamefont {Morrison}}, \ and\
  \bibinfo {author} {\bibfnamefont {E.}~\bibnamefont {Sonnendrücker}},\ }\href
  {\doibase 10.1017/S002237781700040X} {\bibfield  {journal} {\bibinfo
  {journal} {Journal of Plasma Physics}\ }\textbf {\bibinfo {volume} {83}}
  (\bibinfo {year} {2017}),\ 10.1017/S002237781700040X}\BibitemShut {NoStop}%
\bibitem [{\citenamefont {Frankel}(2012)}]{Frankel_2012}%
  \BibitemOpen
  \bibfield  {author} {\bibinfo {author} {\bibfnamefont {T.}~\bibnamefont
  {Frankel}},\ }\href@noop {} {\emph {\bibinfo {title} {The Geometry of
  Physics: An Introduction}}},\ \bibinfo {edition} {3rd}\ ed.\ (\bibinfo
  {publisher} {Cambridge University Press},\ \bibinfo {year}
  {2012})\BibitemShut {NoStop}%
\end{thebibliography}%

\end{document}